\documentclass[aip,jcp,amsmath,amssymb,reprint,nofootinbib]{revtex4-1}

\usepackage{bm}
\usepackage{graphicx}
\usepackage{epsfig}
\usepackage{color}
\usepackage{nicefrac}
\usepackage{mathrsfs}
\usepackage{chemarrow}
\usepackage{textcomp}
\usepackage[all,2cell]{xy}
\usepackage{multirow} 
\usepackage{nicefrac}
\UseTwocells

\def\dbar{{\mathchar'26\mkern-13mu d}}
\newcommand{\bi}{\begin{itemize}}
\newcommand{\ei}{\end{itemize}}

\newcommand{\be}{\begin{equation}}
\newcommand{\ee}{\end{equation}}
\newcommand{\bea}{\begin{eqnarray}}
\newcommand{\eea}{\end{eqnarray}}
\newcommand{\bes}{\begin{subequations}\bea}
\newcommand{\ees}{\eea\end{subequations}}
\newcommand{\ba}{\begin{array}}
\newcommand{\ea}{\end{array}}
 
\newcommand{\bs}[1] {\boldsymbol{#1}}
\newcommand{\TO}{\ar@<.2ex>@{-^{>}}}
\newcommand{\FROM}{\ar@<-.2ex>@{_{<}-}}

\newcommand{\WATER}{\mathrm{H}_2\mathrm{O}}

\newcommand{\nY}{\nabla^{Y}}
\newcommand{\nX}{\nabla^{X}}
\newcommand{\onabla}{\overline{\nabla}}
\newcommand{\onX}{\overline{\nabla}^{X}}
\newcommand{\onY}{\overline{\nabla}^{Y}}

\DeclareRobustCommand{\gobblefour}[4]{}
\newcommand*{\SkipTocEntry}{\addtocontents{toc}{\gobblefour}}

\begin{document}
 
%\preprint{APS/123-QED}

\title{Irreversible thermodynamics of open chemical networks I:\\ Emergent cycles and broken conservation laws}
 
\author{Matteo Polettini}
 \email{matteo.polettini@uni.lu}
\affiliation{Complex Systems and Statistical Mechanics, University of Luxembourg, Campus
Limpertsberg, 162a avenue de la Fa\"iencerie, L-1511 Luxembourg (G. D. Luxembourg)} 

\author{Massimiliano Esposito}
\affiliation{Complex Systems and Statistical Mechanics, University of Luxembourg, Campus
Limpertsberg, 162a avenue de la Fa\"iencerie, L-1511 Luxembourg (G. D. Luxembourg)}

\date{\today}

\begin{abstract}
In this and a companion paper we outline a general framework for the thermodynamic description of open chemical reaction networks, with special regard to metabolic networks regulating cellular physiology and biochemical functions. We first introduce closed networks ``in a box'', whose thermodynamics is subjected to strict physical constraints: the mass-action law, elementarity of processes, and detailed balance. We further digress on  the role of solvents and on the seemingly unacknowledged property of network independence of free energy landscapes. We then open the system by assuming that the concentrations of certain substrate species (the {\it chemostats}) are fixed, whether because promptly regulated by the environment via contact with reservoirs, or because nearly constant in a time window. As a result, the system is driven out of equilibrium. A rich algebraic and topological structure ensues in the network of internal species: Emergent irreversible cycles are associated to nonvanishing affinities, whose symmetries are dictated by the breakage of conservation laws. {\color{black}These central results are resumed in the relation $a + b = s^Y$ between the number of fundamental affinities $a$, that of broken conservation laws $b$ and the number of chemostats $s^Y$.} We decompose the steady state entropy production rate in terms of fundamental fluxes and affinities in the spirit of Schnakenberg's theory of network thermodynamics, paving the way for the forthcoming treatment of the linear regime, of efficiency and tight coupling, of free energy transduction and of thermodynamic constraints for network reconstruction. 
\end{abstract}

% \pacs{05.70.Ln,87.16.Yc}

% 05.70.Ln Nonequilibrium and irreversible thermodynamics
% 87.16.Yc Regulatory genetic and chemical networks

\maketitle

%\tableofcontents

\section{Introduction\label{intro}}

\subsection{Foreword} 

Ever since Schr\"odinger's visionary essay\cite{schroedinger}  {\it What is life?}, gene expression, emergence of structure, and entropic throughput have been tightly intertwined. Information, self-organization, thermodynamics: All three threads of this eternal golden braid are currently receiving a new wave of interest based on the development of common theoretical grounds. As regards thermodynamics, after being for long a loose collection of phenomenological laws and universal rules, it is now achieving a unified theoretical framework, encompassing systems far from equilibrium and subject to fluctuations\cite{broeck,seifert1,jarzynski}. The modern understanding of nonequilibrium states is particularly effective on networks representing fluxes of matter, charge, heat and more generally of information. A line of inquiry founded in the works of Kirchhoff on electrical circuits has recently approached the complex networks involved in biochemical modeling \cite{turing,hillfree,oster,clarke,ederer,qian,andrieux,seifert}. 

Nevertheless, there still are enormous gaps between the foundations of statistical thermodynamics and the phenomenological modeling of chemical networks. The goal of the present paper, {\it quasi una} review, and of its companion paper to be published soon\cite{paper2}, is to provide an organic thermodynamic picture of a self-regulating network of  chemical reactions internal to a system driven out of equilibrium by the action of certain reservoirs of external chemical substrates (the {\it chemostats}), whose availability is independently administered by the environment through membranes and pores ({\it in vivo} conditions) or is practically constant within the timescales of interest ({\it in vitro} conditions). In many respects our contribution inscribes in an effort by several authors\cite{clarke,qian} to provide the chemical analog of the network theory of macroscopic observables reviewed by Schnakenberg\cite{schnak} in a milestone of nonequilibrium thermodynamics. {\color{black} As we will show, the theory becomes significantly richer when it is generalized from the linear dynamics of populations, regulated by the master equation, to the nonlinear dynamics of chemical species, regulated by mass-action kinetics.}

The paradigmatic case that this theory aspires to address are large metabolic networks \cite{soh}, envisaged as complex thermodynamic machines. Metabolic networks regulating cellular physiology, biochemical functions and their correlation to gene expression are being sequenced in greater and greater detail by various genome projects, the first example being the {\it Escherichia coli} metabolism, whose stoichiometry has been assembled by Edwards and Palsson\cite{edwards}. They typically include hundreds to thousands pathways that cannot be modeled exhaustively but need to be reconstructed solely from knowledge of network data. In this effort, compliance to the laws of thermodynamics has been addressed by various authors\cite{qian,TMFA,demartino}. 
% in the context of Energy Balance Analysis\cite{qian} and of Thermodynamic-Based Metabolic Flux Analysis\cite{TMFA}.
In fact, metabolism is the archetype of a nonequilibrium process: By feeding and expelling chemical species (and carriers of radiation), internal processes constantly produce entropy to maintain fluxes of energy and substances. Our ultimate goal is to provide tools for the quantification of the performed work, the dissipated heat, the efficiency of processes and the response to perturbations.
 
Examples of chemostats in biophysical systems are numerous, depending on where the boundary is traced between the system and the environment. Let us mention a few. The usual solvent in biochemical reactions is water\cite{ederer}, typically supplied through osmosis. For this reason it will be convenient to think of water as the ``ground''  chemostat. Chemiosmosis is one among many mechanisms regulating the availability of $\mathrm{H}^+$ ions transported by the proton-motive force across membranes. All physiological solutions are buffered, that is, held in a narrow pH window. Homeostasis in humans and animals maintains fixed concentrations of glucose in blood. While enzymes are typically internal to the cell, their organic or inorganic cofactors such as metal ions might be replenished by the environment. Molecules involved in the supply of tokens of energy, the ubiquitous adenosine tri-, di- and mono- phosphate and inorganic phosphate, and in particular the concentration ratio ATP/ADP determining the energy available to cells to perform biosynthesis, molecular pumping, movement, signal transduction and information processing\cite{qian2} are assumed to be tightly regulated. Other examples of chemostats might include oxygen and carbon dioxide in respiration; nutrients and biomass in metabolism; models of the E. Coli metabolism includes glucose, ammonium, sulfate, oxygen and phosphate as substrates \cite{hoppe}; eventually, light acts as a chemostat responsible for complex behavior such as bistability \cite{ross}, though we will not consider interaction with radiation here. 

Biochemical modeling is usually ``top-down'', aiming at reconstructing the mechanisms that might underlie an observable macroscopic behavior. Often, precisely because their concentrations are constant, chemostats are not explicitly incorporated in the description. Similarly, the effect of enzymes is usually coarse-grained, leading to peculiar kinetic laws\cite{schuster}. On the contrary, our approach is ``bottom-up''. A main theme in our work is that not only do chemostats affect the thermodynamic understanding of chemical networks, but indeed they are the species of observational interest in many respects. Also, while enzymes notoriously do not alter the free energy landscape, but rather are responsible of regulating the reactions' velocities, it will follow from our analysis that all conserved quantities (such as enzymatic free and substrate-bound states) have a thermodynamic role when networked.

We therefore propose a theory of nonequilibrium thermodynamics of chemostatted chemical networks which builds from fundamental physical requirements that elementary reactions should abide by. While the first half of this paper is mainly devoted to characterizing the physical requirements, in the second part we derive the novel results. In particular we provide a cycle decomposition of the entropy production, distinguishing between detailed balanced cycles and {\it emergent} cycles that carry nonvanishing chemical affinities. We are thus able to express the steady state entropy production in terms of the chemostats' chemical potentials and currents. We further show that chemostatting breaks conservation laws when a balance of conserved quantities (such as mass) across the system's boundary is induced. Each such broken conservation law corresponds to symmetries of the affinities, hence it describes the internal redundancy of the nonequilibrium currents. Finally we show that solvents play the role of ``ground'' chemostats, and provide a fundamental relation between the number of chemical affinities, of broken conservation laws and of chemostats.

\subsection{Plan and notation}

After commenting in the next paragraph on the choice of chemostats as the fundamental thermodynamic forces, we illustrate our main findings with a simple example in Sec.\;\ref{example}. In Sec.\;\ref{gen} we introduce Chemical Networks (CN) as they are described in the biochemistry/applied mathematics literature. In Sec.\;\ref{close} we impose further constraints on closed CNs and introduce the thermodynamic machinery of chemical potentials. In Sec.\;\ref{open} we open up the network. We draw partial conclusions in Sec.\;\ref{conclu}, postponing further discussion to the follow-up paper. {\color{black}We provide another example in Appendix \ref{appex}.}

Dynamic and thermodynamic perspectives on CNs employ different languages; we tried to accommodate both. % The reader not confident with thermodynamic potentials can altogether skip all formulas containing $\bs{\mu}$ and $\Delta \bs{G}$ from Sec.\;\ref{gen} without great loss.
Vectors are bold, $\bs{w}$; their entries are labelled with greek indexes, $w_\sigma$. The scalar product between vectors of the same dimension is $\bs{v}\cdot \bs{w}$. Matrix transposition is $\overline{M}$. As is customary in the CN literature, analytic functions of vectors are defined component-wise:
{\color{black}
\be
\left(\ln \bs{w}\right)_\sigma = \ln w_\sigma, \quad \left(\frac{\bs{w}}{\bs{w}'} \right)_\sigma  =  \frac{w_\sigma}{ w'_\sigma}, \quad \ldots
\ee
}
The ``dot power'' of a vector by a vector of the same length will imply the scalar product in the following way:
\be
\bs{w}^{\, \cdot \bs{w}'} = e^{ \bs{w}'  \cdot \ln \bs{w}} = \prod_\sigma w_\sigma^{w'_\sigma}. \qquad 
\ee
Stoichiometric matrices are denoted $\nabla$.

We will make use of several indices, whose range and meaning is here reported fo later reference:
\bi
\item[] $\sigma = 1,\ldots, s$ for species ($s = \#$ rows $\nabla$);
\item[] $\rho = 1, \ldots, r$ for reactions ($r = \#$ columns $\nabla$);
\item[] $\gamma = 1, \ldots,c$ for cycles ($c = \dim \ker \nabla$);
\item[] $\lambda = 1,\ldots,\ell$ for conservation laws ($\ell = \dim \ker \overline{\nabla}$);
\item[] $\alpha = 1, \ldots ,a$ for affinities ($a = \dim \ker \nabla^X - c$);
\item[] $\beta = 1,\ldots ,b$ for broken conservation laws\\ $~\qquad \qquad \qquad \qquad \qquad~$ ($b = \ell - \dim \ker \overline{\nabla}^X$).
\ei
Finally, we use the abbreviations: Chemical Network (CN), Closed CN (CCN), Open CN (OCN), Kirchhoff's Current Law (KCL), Kirchhoff's Loop Law (KLL), Entopy Production Rate (EPR), Flux Balance Analysis (FBA), Energy Balance Analysis (EBA).

\subsection{Why chemostats?}

Flux Balance Analysis (FBA)\cite{bonarius} and Metabolic Flux Analysis\cite{stepha} are methods of reconstruction of the steady state of a CN based on Kirchhoff's Current Law (KCL). Such constraints are complemented by optimization techniques and experimental data. Energy balance analysis (EBA) \cite{qian} advances these methods by implementing Kirchhoff's Loop Law (KLL) to avoid infeasible thermodynamic cycles leading to violations of the second law of thermodynamics. In these approaches, nonequilibrium steady states are driven by fixed fluxes along boundary reactions. The stoichiometric matrix is parted as
\be
\nabla = \left( \ba{c|c} \multirow{2}{*}{int. reactions} &  \multirow{2}{*}{boundary reactions}     \\  ~ \ea \right),
\ee
supposing no reactant is completely externalized. In our approach we part it as
\be
\nabla = \left( \ba{c}  \mathrm{int. ~ reactants}  \\ \hline  \mathrm{ext. ~ reactants}  \ea \right),
\ee
supposing no reaction is completely externalized. As we will comment on in the follow-up paper, as far as steady current configurations are considered, the two approaches yield equivalent results, since such currents can be effectively obtained by degradation of a chemostat\cite{horn}. {\color{black}We emphasize that many authors considered the two different approaches to opening networks. see e.g.\cite{qianclamp}}. Let us here advance some motivations why it is useful to move the focus on chemostats for thermodynamic modeling.

First, beside steady states, in thermodynamics one is also interested in the process of relaxation. In this respect, fixed currents are problematic as they are incompatible with mass-action kinetics. A fixed incoming stream of particles can be modeled by the reaction
$\emptyset \to Z$ that creates a molecule of $Z$ at fixed rate\cite{foot1}. However, the inverse reaction $\emptyset \gets Z \label{eq:empty2}$  provides a term proportional to the concentration of $Z$, and there is no way to stabilize a fixed negative current in the process of relaxation to the steady state. % That is, mass-action kinetics only accounts for  fixed incoming currents; outward currents vary depending on the reactor's content
On the other hand, non-mass-action kinetics with fixed external currents is inconsistent with the preparation of certain initial states, e.g. an empty reactor. Moreover, the vast literature on CNs is invariably based on mass-action, in particular as regards the existence, uniqueness and stability of steady states. Further arguments in favor of mass-action will be reported later.

Second, in linear nonequilibrium thermodynamics \cite{prigo} the response of systems to external perturbations is usually formulated as the re-organization of currents after a modification of their conjugate forces, i.e. the chemostats' chemical potential differences. In this respect our theory fits this paradigm by construction.

Finally, and most importantly, the results presented in this paper are meant to be applicable to the stochastic thermodynamics associated to the chemical master equation \cite{nicolis,rossbook,seifert,bazzani}, where one typically deals with force constraints (encoded in the rates) rather than with current constraints. Chemostats play a central role in the derivation of advanced results in the stochastic theory of chemical reactions, such as the Fluctuation Theorem for the currents and the Green-Kubo relations\cite{andrieux}.

\section{Chemostatting: An example\label{example}}

Consider the CN
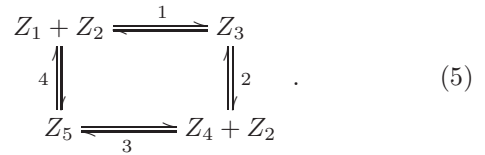
\be
\ba{c}\xymatrix{
Z_1 + Z_2 \FROM[r]\TO^{~~~~1}[r] & Z_3 \FROM[d]\TO^{2}[d]  \\
Z_5  \FROM[u]\TO^4[u] & Z_4 + Z_2 \FROM[l]\TO^{3~~~~}[l]}
\ea. \label{eq:excn} \ee
The system is closed, that is, no fluxes of matter are allowed in or out of the reactor. Hence, all species' concentrations $[Z_\sigma]$ vary as reactions occur. Though, they are not independent. Letting $m_\sigma \geq 0$ be the molar masses of species, necessarily satisfying $m_1 = m_4$, $m_3 = m_5= m_1+m_2$\cite{foot2}, then the total mass per unit volume
\be
L_1 = \sum_\sigma m_\sigma  [Z_\sigma]
\ee
is conserved. The combination
\be
L_2 = [Z_1]+[Z_3]+[Z_4]+[Z_5]
\ee
is also conserved, a symptom that $Z_2$ acts like an enzyme\cite{foot3}.

Let us consider the cyclic transformation
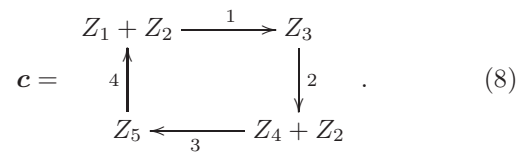
\be
\bs{c} = \ba{c} \xymatrix{
Z_1 + Z_2 \ar@{<-}_4[d] \ar@{->}^{~~~~1}[r] & Z_3 \ar@{->}^2[d]  \\
Z_5 \ar@{<-}_{3~~~~}[r] & Z_4 + Z_2}
\ea .\ee
The free energy increase along the first reaction is
\be
\Delta_1 G = \mu_3 - \mu_1 - \mu_2
\ee
where $\mu_{\sigma}$ are chemical potentials, and so on for other reactions. When the cycle is closed, we obtain a vanishing {\it affinity}:
\be
\mathcal{A}(\bs{c}) = \frac{1}{RT}(\Delta_1 + \Delta_2 + \Delta_3 + \Delta_4) G = 0.
\ee
This is due to the fact that free energy is a state function, reflecting the reversible nature of the system: At equilibrium the average number $\mathcal{J}(\bs{c})$ of completions of the cycle in the clockwise direction equals that  $\mathcal{J}(-\bs{c})$ in the counterclockwise direction\cite{foot4}. 

Let $J_{+ \rho}$  be the number of moles per unit time and unit volume that perform reaction $\rho$ in clockwise direction, $J_{-\rho}$ in counterclockwise direction, and let $J_\rho = J_{+\rho}- J_{-\rho}$ be the net reaction flux. Letting $T$ be the environmental temperature, the entropy production rate (EPR) is defined as the rate of free energy decrease\cite{kondepudi}
\bea
T \Sigma & = &  - \sum_\rho J_\rho \Delta_\rho G \\
& = & \mu_1(J_1 - J_4) + \mu_2(J_1 - J_4 + J_3 - J_2) \nonumber \\ & & + \; \mu_3(J_2 - J_1) + \mu_4(J_3 - J_2) + \mu_5(J_4 - J_3). \nonumber 
\eea
At a steady state concentrations arrange themselves in such a way that fluxes balance each other for each species according to KCL: 
\be J^\ast_4-J^\ast_1 = J^\ast_1 - J^\ast_2 = J^\ast_2 - J^\ast_3 = J^\ast_3-J^\ast_4 = 0. \label{eq:kclclose} \ee
The most general solution to Eq.\,(\ref{eq:kclclose}) is
\be
J^\ast_1 = J^\ast_2 = J^\ast_3 = J^\ast_4 = \mathcal{J}. \label{eq:kir}
\ee
In the setup we have so far specified, the cycle current $\mathcal{J}$ needs not vanish. On the other hand, by direct substitution it can be shown that the steady EPR vanishes, so that no dissipation occurs within the system at the steady state:
\be
\Sigma^\ast = 0.
\ee
Then, one might conclude that it is still possible that currents flow within the system without dissipation. This is clearly incompatible with the laws of thermodynamics, as one would obtain a {\it perpetuum mobile}. Notice that so far no assumption has been made on the kinetics of the system, i.e. on how currents respond to their driving forces, the chemical potential differences. If individual reactions are thermodynamically independent one of another, then the current should always be driven in the same direction as its corresponding chemical potential gradient, so that
\be
- J^\ast_\rho \, \Delta_\rho G \geq 0. \label{eq:EBA}
\ee
Under this requirement necessarily $\mathcal{J} = 0$, and the steady state is an equilibrium $[Z_\sigma]^{eq}$ with vanishing currents, $J^{eq}_\rho = 0$. The above equation is the core of EBA\cite{qian}. Such constraints are inbuilt in mass-action kinetics\cite{foot5}. We point out that Eq.\,(\ref{eq:EBA}) should hold for elementary reactions; non-elementary reactions might allow coupling between different mechanisms that yield negative response of currents to forces (see part II).

We now open the system by chemostatting external species $Y_1=Z_1$ and $Y_4=Z_4$. Every time a molecule of this kind is consumed or produced by a reaction, the environment withdraws or provides one so to keep the chemostats' concentrations fixed. All other varying internal species will be denoted $X$. The open network is depicted by
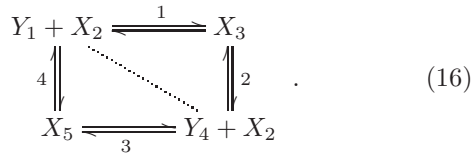
\be
\ba{c}\xymatrix{
Y_1 + X_2 \FROM[r]\TO^{~~~~1}[r] & X_3 \FROM[d]\TO^{2}[d]  \\
X_5  \FROM[u]\TO^4[u] & Y_4 + X_2 \FROM[l]\TO^{3~~~~}[l] \ar@{.}[ul]} \ea. \label{CN:dot}
\ee
If the concentrations of the chemostats are held at their equilibrium values $[Y_1]\equiv [Z_1]^{eq}$, $[Y_4] \equiv [Z_4]^{eq}$, all other species also equilibrate to the same value they attained in the closed network. Otherwise, since there is a source and a sink of substances, we expect the system to move to a nonequilibrium steady state capable of transferring free energy from one reservoir to the other. 

Nonequilibrium systems are characterized by thermodynamic cycles that are typically performed in a preferential direction so to produce a positive amount of entropy. In the first segment of cyclic transformation $\bs{c}$ the environment provides one molecule of $Y_1$ that reacts with $X_2$ to produce $X_3$. Inside the reactor there is no variation of the concentration of $Y_1$. The internal free energy increase along this transformation is then
\be
\Delta_1 G^X = \mu_3 - \mu_2.
\ee
The chemical work needed to provide one molecule to the reactor is given by the external free energy increase
\be
\Delta_1 G^Y = - \mu_1.
\ee
Around the cycle,
\bes
(\Delta_1 + \Delta_2 + \Delta_3 + \Delta_4) G^X & = & 0, \\
(\Delta_1 + \Delta_2 + \Delta_3 + \Delta_4) G^Y & = & 0. 
\ees
Both the internal and the external free energy cycles vanish, hence no free energy is transferred from one reservoir to the other. Given the preamble, this is a bit surprising, as we would expect irreversible cycles.

This is indeed the case, though in a more subtle way. Notice that states $Y_1 +X_2$ and $Y_4 + X_2$ along the cycle are indistinguishable (as highlighted in Eq.\;(\ref{CN:dot}) by a dotted line), since after transitions $1$ and $2$ all concentrations come to coincide again. Then, the true state space is found by lumping together the two identical states:
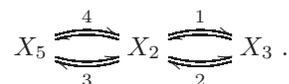
\be
\xymatrix{
X_5 \FROM@/_/[r]_3\TO@/_/[r]   \FROM@/^/[r]\TO@/^/[r]^4 & X_2  \FROM@/_/[r]_2\TO@/_/[r]   \FROM@/^/[r]\TO@/^/[r]^1 & X_3}. \nonumber
\ee
This contracted network allows for two cycles that consume a molecule of $Y_1$ and produce one of $Y_4$,
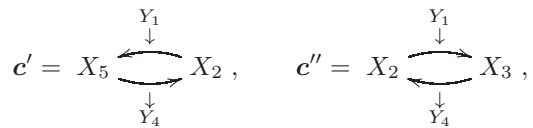
\be
\bs{c}' = \xymatrix{X_5 \ar @{->} @/_/[r]_{\substack{\downarrow \\ Y_4}} \ar @{<-} @/^/[r]^{\substack{Y_1 \\ \downarrow}} & X_2 }, \qquad \bs{c}'' =  \xymatrix{X_2 \ar @{<-} @/_/[r]_{\substack{\downarrow \\ Y_4}} \ar @{->} @/^/[r]^{\substack{Y_1 \\ \downarrow}} & X_3 }, \nonumber
\ee
yielding a net flux of matter from one reservoir to the other. The free energy balance along $\bs{c}''$ is
\be
(\Delta_1 + \Delta_2) G^X = 0, \qquad
(\Delta_1 + \Delta_2) G^Y = \mu_4 - \mu_1,
\ee
and similarly for $\bs{c}'$. Work is performed to displace substances and is degraded into heat; a nonnull affinity $\mathcal{A}(\bs{c}') =  \mathcal{A}(\bs{c}'') = (\mu_4 - \mu_1)/RT$ characterizes these cycles. Notice that while $G^X$ is still a state function, on the contracted network $G^Y$ is not a state function: In fact, it is multi-valued at the lumped states, and its increase depends on the path between two states.
 
We can further define an internal and an external contributions to the EPR,
\be
T\Sigma^{X,Y} = -\sum_\rho J_\rho \; \Delta_\rho G^{X,Y}.\label{eq:ep}
\ee
Explicitly we get
\bea
T\Sigma^X & = & \mu_2(J_1 - J_4 + J_3 - J_2) + \mu_3(J_2 - J_1) + \mu_5(J_4-J_3) \nonumber \\
T\Sigma^Y & = & \mu_1 (J_1 - J_4) + \mu_4 (J_3-J_2). \label{eq:exexex}
\eea
At the steady state KCL holds:
\be
J^\ast_4- J^\ast_1 = J^\ast_3-J^\ast_4 = J^\ast_1 + J^\ast_3 - J^\ast_2 - J^\ast_4 = 0.
\ee
Quite crucially, since two states are lumped into one, one such law is lost with respect to Eq.\;(\ref{eq:kclclose}). This implies that external steady EPR does not vanish: 
\be
T{\Sigma^X}^\ast = 0, \qquad
T{\Sigma^Y}^\ast = (\mu_1-\mu_4)  \left(\frac{\dbar [Y_1]}{\dbar t}\right)^\ast,
\ee
where  $\,\dbar [Y_1] / \, \dbar t = J_1 - J_4$ is the rate at which chemostat $Y_1$ is injected into the system. The slash derivative is a notation that in the later development will signify that these rates are {\it not} exact time derivatives.

A few considerations regarding cycles and conservation laws. In the open network the internal mass $\sum_{\sigma = 2,3,5} m_\sigma [X_\sigma]$ is not conserved since, for example, reaction 1 increases the system's mass by $m_1$. Yet there still survives one conservation law:
\be
L' = [X_2]+[X_3]+[X_5].
\ee
Mass conservation is already broken by chemostatting one species only, in which case we would obtain no new cycle. In fact there exists a precise relation between number of chemostats, of broken conservation laws and of emergent irreversible cycles. The effect of the broken mass law is still visible in the open network, as it implies that $(\, \dbar [Y_1] / \, \dbar t)^\ast   = - (\, \dbar [Y_4] / \, \dbar t )^\ast $, i.e. at steady states the injected current of chemostat $Y_1$ equals the ejected current of chemostat $Y_4$. Broken conservation laws also encode symmetries: The equilibrium condition $\mu_4 = \mu_1$ implies that the relative concentration $[Y_1]/[Y_4]$ must attain its equilibrium value, but individual concentrations can be proportionally increased without altering the affinity. 

Let us finally impose the thermodynamic constraints in Eq.\;(\ref{eq:EBA}) to the OCN curents, which can be expressed in terms of two boundary currents:
\be
J_1^\ast = J_2^\ast = \mathcal{J}_1, \qquad J_3^\ast = J_4^\ast = \mathcal{J}_2.
\ee 
One obtains the inequalities
\bea
\mathcal{J}_1 \, \Delta_1 G \leq  0; \quad
\mathcal{J}_1 \, \Delta_2 G \leq 0; \quad
\mathcal{J}_2 \, \Delta_3 G \leq 0; \nonumber \\
- \mathcal{J}_2 \, (\Delta_1 G + \Delta_2 G  + \Delta_3 G) \leq 0. \qquad
\eea
After some manipulations one finds
\be
\frac{\mathcal{J}_1}{\mathcal{J}_2} \leq 0,
\ee
which implies that, {\color{black} assuming $\mu_1 > \mu_4$}, the two cycle currents around $\bs{c}_1$ and $\bs{c}_2$ should wind in the direction that consumes $Y_1$ and produces $Y_4$:
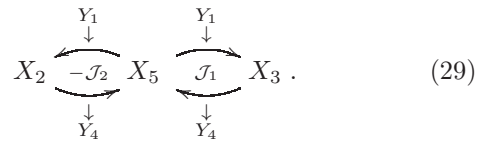
\be
\xymatrix{ X_2  \rtwocell_{\substack{\vspace{0.3cm} \\ \downarrow \\ Y_4}}^{\substack{Y_1 \\ \downarrow \vspace{0.3cm}}} {`-\mathcal{J}_2} & X_5 \ & X_3  \ltwocell^{\substack{\vspace{0.3cm} \\ \downarrow \\ Y_4}}_{\substack{Y_1 \\ \downarrow \vspace{0.3cm}}}{'\mathcal{J}_1}}. \label{eq:cycless}
\ee
This echoes the principle that like causes produce like effects. To conclude, let us resume the key takeaways of this introductory example:
\bi
\item[--] A CCN equilibrates. An OCN is obtained by externally maintaining the concentrations of certain species, the chemostats.
\item[--] For every chemostat, either a conservation law is broken or a new cycle emerges.
\item[--] While free energy is an equilibrium state function on the CCN, it is not on networks with lumped states since its circuitation along emergent cycles usually produces a nonvanishing affinity, a marker of nonequilibrium behavior.
\item[--] At steady states the EPR can be expressed in terms of the chemostats' chemical potential differences and currents only.
\item[--] Broken conservation laws imply symmetries of the affinities with respect to chemostat concentrations and linear relationships between inflows of chemostats.
\item[--] Thermodynamic feasibility of network reconstruction can be implemented by requiring that currents satisfying KCLs in the OCN are compatible with existence of a free energy landscape on the CCN (to be discussed in the follow-up paper).
\ei
 
\section{\label{gen}Chemical Networks}

\subsection{Setup}

We consider a reactor with volume $V$ where species $\bs{Z} = (Z_\sigma)$ engage into chemical reactions $\rho$. We will first consider closed systems for which no exchange of matter with the environment occurs. Nevertheless, the reactor is not isolated: An energy-momentum trade-off thermalizes the individual species' distribution to a Maxwellian at environmental temperature $T$. The solution is well-stirred, i.e. molecules are homogeneously distributed in space. Their number is of the order of Avogadro's number, so that fluctuations are irrelevant\cite{foot6}. Also, we assume the mixture to be ideal (satisfying Raoult's law)\cite{foot7}.

The concentration of species $Z_\sigma$ is $[Z_\sigma]$, with dimension of moles per liter. They appear as reactants or products in different combinations, called {\it complexes}. We assign an arbitrary orientation to each reaction $+\rho$, from a complex of reactants to a complex of products. This is just a notational convention, insofar as every reactant occurs as a product along the inverse reaction $-\rho$. The $s \times r$ matrices $\nabla_{\!+} = (\nabla_{\!\sigma,+\rho})$ and $\nabla_{\!-} = (\nabla_{\! \sigma, -\rho})$ collect the stoichiometric coefficients of the reactants and of the products respectively (note that species ``to the left'' are marked plus and species ``to the right'' are marked minus). We denote $\nabla_{\! \pm\rho}$ their column vectors, describing the stoichiometry of a given reaction. The stoichiometric matrix is
\be
\nabla = \nabla_{\!-} - \nabla_{\!+} \; \stackrel{\mathrm{e.g.}}{=} \; \stackrel{{\normalsize r}}{\overbrace{\left(\ba{cccc} -1 & 0 & 0 & 1 \\ -1 & 1 & -1 & 1 \\ 1 & -1 & 0 & 0 \\ 0 & 1 & -1 & 0 \\ 0 & 0 & 1 & -1 \ea\right)}} \left.\ba{c} \, \\ \, \\ \, \\ \, \\ \, \ea \hspace{-0.3cm} \right\} \text{\,\footnotesize{\emph{s}}}. \label{eq:stoichiometric}
\ee
Entry $\nabla_{\!\sigma,\rho}$ yields the net number of molecules of species $Z_\sigma$ produced (consumed) in reaction $+\rho$ ($-\rho$).

\subsection{Mass-action kinetics}

The directed flux $J_{\pm \rho}$ is the rate of change of the concentrations of the species partaking to reaction $\pm \rho$. We will assume the law of mass-action: The directed flux is proportional to the probability of an encounter between the reactants, which on the other hand is proportional to the product of their concentrations (for large numbers of particles):
\be
\bs{J}_\pm = \bs{K}_{\pm} [\bs{Z}]^{\, \cdot \nabla_\pm}. \label{eq:adim}
\ee
The nonnegative rate constants $K_{\pm \rho}$ depend on the microphysics of the individual reaction; a reaction is reversible if both $K_{\pm\rho} > 0$. Rate constants have different physical dimensions depending on the stoichiometry of the reactants.

The (net) flux, or current, is the difference between the directed flux along $+\rho$ and that along $-\rho$:
\be
\bs{J} = \bs{J}_+ -  \bs{J}_- .
\ee
The mass-action kinetic equations finally read 
\be
\frac{d}{dt} [\bs{Z}] = \nabla \bs{J}, \label{eq:cont}
\ee
revealing the rationale for the somewhat unconventional symbol $\nabla$: It acts as a divergence of a current, making Eq.\;(\ref{eq:cont}) into a continuity equation.  However, unlike the discretized divergence operator in algebraic topology, $(1,1,\ldots,1)$  needs not be a left null vector of $\nabla$. As a consequence the total concentration is not necessarily constant, leading to the definition of the total rate 
\be \dot{\mathcal{N}} = \frac{d}{dt}  \ln [Z]
\label{eq:contev}, \ee
where $[Z]= \sum_\sigma [Z_\sigma]$ is the total concentration of the mixture. The total rate vanishes when each reaction preserves the number of reactants.

{\color{black}
For the example in Sec.\,\ref{example} we have the net currents
\bea
J_{1} & = & K_{+1} [Z_1][Z_2] - K_{-1} [Z_3] \nonumber \\
J_{2} & = & K_{+2} [Z_3] - K_{-2} [Z_2][Z_4] \nonumber \\
J_{3} & = & K_{+3} [Z_2][Z_4] - K_{-3} [Z_5] \nonumber \\
J_{4} & = & K_{+4} [Z_5] - K_{-4} [Z_1][Z_2] 
\eea
and the mass-action kinetic equations are given by
\bea \label{eq:exmak}
\tfrac{d}{dt} [Z_1] & = & J_{4} - J_{1} \nonumber \\
\tfrac{d}{dt} [Z_2] & = & J_{4} - J_{1} + J_2 - J_3 \nonumber \\
\tfrac{d}{dt} [Z_3] & = & J_{1} - J_{2} \nonumber \\
\tfrac{d}{dt} [Z_4] & = & J_{2} - J_{3} \nonumber \\
\tfrac{d}{dt} [Z_5] & = & J_{3} - J_{4}
\eea
and it can be easily verified that the total concentration is not conserved.
}

\subsection{Chemical networks and their representations}

All of the above defines a Chemical Network (CN), which will be concisely depicted by the set of stoichiometric equations
\be
\xymatrix{
\nabla_{\!+} \bs{Z}  ~ \FROM[r]_{\bs{K}_-}\TO^{\bs{K}_+}[r] & ~ \nabla_{\!-} \bs{Z}}.
\ee

Merging all identical complexes of reactants, one obtains a graphical representation of the CN as a directed graph between complexes, e.g. Eq.\,(\ref{eq:excn}). There exists an alternative representation of a CN as an hypergraph, with individual species as vertices connected by reaction links that connect many species to many species\cite{klamt}. {\color{black} Our example is represented by the following hypergraph:
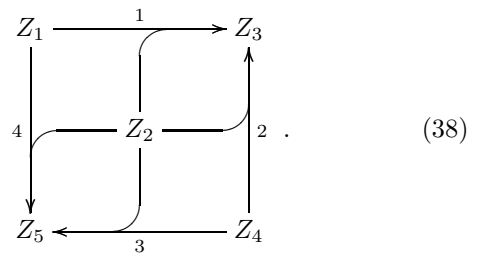
\be
\ba{c}\xymatrix{
Z_1  \ar @{->}_4[dd]  \ar @{->}^1[rr] &     & Z_3 \\
    & Z_2 \ar `d[d] [dl] \ar `u[u] [ur]  \ar 
`r[ur][ur]  \ar 
`l[dl][dl]  &     \\
Z_5 &     & Z_4 \ar @{->}_2[uu]  \ar @{->}^3[ll] }
\ea.
\ee
Here, overlapping arrows depict one unique reaction. We will return to this in \S\;\ref{network}.}

\subsection{Extensive and intensive observables}

The above observables, currents and concentrations, are {\it extensive} in the sense that they scale with the total concentration of the mixture. For thermodynamic modeling it will be convenient to introduce {\it intensive} quantities. The adimensional molar fractions are given by
\be
\bs{z} = [\bs{Z}]/[Z], 
\ee
which are normalized, $\sum_\sigma z_\sigma = 1$. We further define scaled rate constants $k_{\pm \rho} =  [Z]^{\sum_{\sigma}\nabla_{\sigma,\pm \rho}-1}  K_{\pm \rho} $, scaled currents
$\bs{j}_\pm =  \bs{k}_{\pm} \bs{z}^{\,\cdot \nabla_\pm} = [Z]^{-1} \bs{J}_\pm$ and $\bs{j} = \bs{j}_+ - \bs{j}_-$,
with physical dimension of an inverse time. The evolution equation for the molar fractions reads
\be
\frac{d\bs{z}}{dt} =  \nabla \bs{j} - \dot{\mathcal{N}} \bs{z},
\ee
and summing over $\sigma$ one can express the total rate in terms of the scaled currents as $\dot{\mathcal{N}} = \sum_\sigma \nabla_{\!\sigma} \bs{j}$. 

\subsection{Vector spaces of the stoichiometric matrix}

The stoichiometric matrix encodes all relevant information about the topology of a CN \cite{palsson1}.

Every linear operator is characterized by four fundamental vector spaces: The span of its column vectors (image), which is orthogonal to the left-null space (cokernel), the span of its row vectors (coimage), which is orthogonal to the right null space (kernel).

Concerning the stoichiometric matrix $\nabla$:

\begin{itemize}
\item[--] The $\ell$-dimensional cokernel $\mathscr{L}$ is the vector space of  {\it conservation laws}, known as {\it metabolic pools} in the biochemistry literature\cite{palsson1}. Letting $\bs{\ell} \in \mathscr{L}$, i.e. $\onabla \bs{\ell} = 0$, the quantity $L = \bs{\ell} \cdot [\bs{Z}]$ is conserved:
\be
\frac{d L}{dt} = \bs{\ell} \cdot \nabla \bs{J} = 0.
\ee
An enzyme might be defined as a conservation law with all $0$'s and $1$'s, respectively denoting its  bound and free states. Notice that nonlinear equations of motion might admit further nonlinear constants of motion (e.g. the Lotka-Volterra model has no conservation laws but it allows a constant of motion). 
\item[--] The image identifies {\it stoichiometric subspaces} in the space of concentrations where the dynamics is constrained, for given initial conditions. Each stoichiometric subspace is labelled by the value of $\ell$ independent conserved quantities. The dimension of the stoichiometric subspaces is $r - \ell$.
\item[--] The kernel $\mathscr{C}$ is the space of {\it (hyper)cycles}, to which we dedicate a section below. We let $c = \dim \mathscr{C}$.
\item[--] The row space tells in which amounts a molecule is consumed or produced by each reaction.
\end{itemize}
Letting the rank $\mathrm{rk}\, \nabla $ be the number of independent columns and rows of $\nabla$, the rank-nullity theorem in linear algebra states that
\be
\mathrm{rk}\, \nabla = s - \ell = r - c. \label{eq:ranknull}
\ee
We further suppose that neither the row space nor the column space include null vectors (respectively molecules that appear on both sides of all reactions with the same stoichiometry, and reactions between species that do not belong to the network). This would be the case e.g. of an enzyme whose elementary reactions of binding, isomerization and dissociation are not discerned; models of coarse-grained action of an enzyme usually display non-mass-action dynamics\cite{schuster}.

In our example, the stoichiometric matrix in Eq.\;(\ref{eq:stoichiometric}) admits two left and one right null vectors:
\be
\bs{\ell}_1 = \left(\ba{c} 1 \\ 1 \\ 2 \\ 1 \\ 2 \ea\right), \quad \bs{\ell}_2 = \left(\ba{c} 1 \\ 0 \\ 1 \\ 1 \\ 1 \ea\right), 	\quad \bs{c} =  \left(\ba{c} 1 \\ 1 \\ 1 \\ 1 \ea\right).
\ee
The rank-nullity identity reads $3 = 5 - 2 = 4 - 1$. {\color{black}Correspondingly we have two conservation laws
\bes
\tfrac{d}{dt} ([Z_1] + [Z_2] + 2 [Z_3] + [Z_4] + 2[Z_5]) & = & 0 \\
\tfrac{d}{dt} ([Z_1] + [Z_3] + [Z_4] + [Z_5]) & = & 0
\ees
as can be immediatly verified by applying Eqs.\,(\ref{eq:exmak}).
}

\subsection{Steady states and cycle currents}

The cycle space conveys crucial information regarding the nonequilibrium nature of steady states. Cycles are invoked when considering the steady solutions $[\bs{Z}]^\ast$ of the kinetic equation, which satisfy KCL
\be
\nabla \bs{J}^\ast = 0.
\label{eq:steadyvel}
\ee
It is important to note that Eq.\;(\ref{eq:steadyvel}) will usually admit one (or several or no) steady states in each stoichiometric subspace where the dynamics is constrained. Then, steady states will be parametrized by a complete set $\bs{L} = (L_1,\ldots,L_\ell)$ of conserved quantities, $[\bs{Z}]^\ast_{\bs{L}}$. As the system is prepared at an initial state with concentrations $[\bs{Z}](0)$, one can read off the values of the conserved quantities via $\bs{L} = \bs{\ell}\cdot [\bs{Z}](0)$.

A steady solution is said to be an {\it equilibrium} when all currents vanish
\be
\bs{J}^{eq} = 0.
\label{eq:equivel}
\ee
Otherwise it is called a nonequilibrium steady state. Notice that, quite importantly, in this paper we reserve the word ``equilibrium'' to steady states with vanishing currents, while in the mathematically-oriented literature by ``equilibrium'' one usually simply means ``steady state''. Since the kinetic equations are nonlinear, the issue of existence of multiple solutions to Eq.\,(\ref{eq:steadyvel}), or of periodic or chaotic attractors, global and local stability etc. are advanced problems in dynamical systems/algebraic varieties\cite{toric}, that we are not concerned with. In the following, we are rather interested in the thermodynamic characterization of such locally stable steady states, if and when they exist. {\color{black} See Sec.\,(\ref{detbal}) for the determination of equilibrium states of the above example.}

Equation (\ref{eq:steadyvel}) implies that the steady state fluxes $\bs{J}^\ast$ belong to the kernel of the stoichiometric matrix. As a trivial case, if the kernel is empty, then steady fluxes vanish. This is a particular instance of a CN that only allows equilibrium steady states. In general, the kernel of the stoichiometric matrix is not empty. Let $\bs{c}_\gamma$ be basis vectors for $\mathscr{C}$. Then by Eq.\,(\ref{eq:steadyvel}) there exist $c$ cycle currents $\mathcal{J}_\gamma$ such that 
\be
\bs{J}^\ast = \sum_\gamma \mathcal{J}_\gamma \bs{c}_\gamma,
\ee
{\color{black}corresponding to Eq.\,(\ref{eq:kir}) in our example.} Cycle currents are suitable combinations of independent fluxes that suffice to fully describe the steady state. They can be directly calculated in terms of the steady currents as follows. Letting $(\bs{c}^\dagger_\gamma)$ be a set of vectors dual to the cycle vectors in the sense that
$\bs{c}^\dagger_\gamma \cdot \bs{c}_{\gamma'} = \delta_{\gamma \gamma'}$, then
\be
\mathcal{J}_\gamma = \bs{c}^\dagger_\gamma \cdot \bs{J}^\ast. 
\ee
Such a set of covectors always exists, and it is not unique. {\color{black} One such vector in the example is $\overline{\bs{c}}^\dagger = (1,0,0,0)$.}

\subsection{\label{network}Algebraic vs. network cycle analysis}

While in this paper we pursuit a purely algebraic characterization of network thermodynamics, it is worth mentioning that deep questions are related to the network representation of CNs (this paragraph can be safely skipped in view of the forthcoming discussion).

Cycle vectors depict successions of transformations such that the initial and final concentrations coincide. However, cycles might not be visualized as actual cycles in the graphical representation of a CN. Consider
{\color{black}
\be
\ba{c}
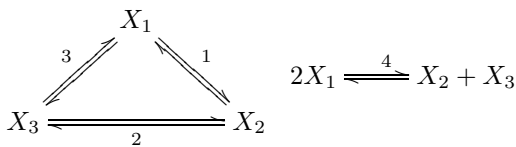
\xymatrix{
& X_1 \FROM[dr]\TO^{1}[dr]  \\
X_3 \FROM[ur]\TO^3[ur] & & X_2 \FROM[ll]\TO^2[ll]}\ea 
\xymatrix{
2X_1& X_2 + X_3 \FROM_{\!\!4}[l]\TO[l]} \nonumber
\ee
with stoichiometric matrix
\be
\nabla = \left(\ba{cccc} -1 & 0 & 1 & -2 \\ 1 & -1 & 0 & 1 \\ 0 & 1 & -1 & 1 \ea \right).
\ee
It admits two linearly independent basis cycles: 
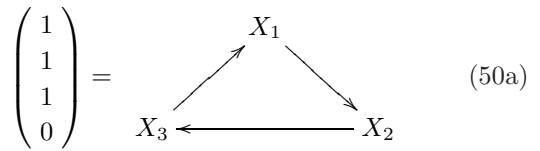
\bes
\left(\ba{c} 1 \\ 1 \\ 1 \\ 0 \ea \right)
 & =  & \ba{c}\xymatrix{ & \ar @{->}[dr] 
X_1   \\
X_3 \ar @{->}[ur] &  &  \ar @{->}[ll]  X_2 }	\ea \\
\left(\ba{c} -1 \\ 0 \\ 1 \\ 1 \ea \right)  & = & \ba{c}
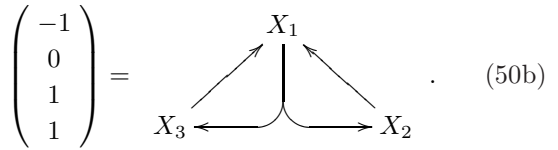
\xymatrix{ & \ar @{<-}[dr] 
X_1  \ar `d[d] [dl] 
 \ar  `d[dr] [dr]
& \\
X_3 \ar @{->}[ur] & & X_2 } \ea .
% \ba{c}\xymatrix{ \ar @{<-}[dr] 
% X_1  \\
% X_3  \ar @{->}[u]  & X_2 } \ea 
% \xymatrix{
% 2X_1  \ar @{->}[r] & X_2 + X_3} . \nonumber
\ees
The first is a closed path in the graph-theoretical sense. 
More subtly, the second is an  hyper-cycle. 
The difference between a graph cycle and an hypercycle is that the first involves transformations that always preserve the complexes, while an hypercycle eventually dismembers complexes to employ their molecules in other reactions, as is the case for the complex $2X_1$ whose individual molecules $X_1$ are employed in reactions $1$ and $3$. This subtle difference can be spotted by replacing the detailed information about species by abstract symbols for complexes, e.g.
\be
\ba{c}
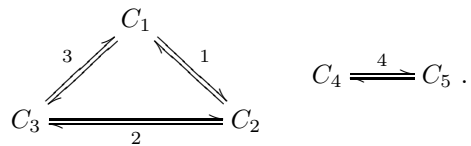
\xymatrix{
& C_1 \FROM[dr]\TO^{1}[dr]  \\
C_3 \FROM[ur]\TO^3[ur] & & C_2 \FROM[ll]\TO^2[ll]}\ea \quad
\xymatrix{
C_4 & C_5 \FROM_4[l]\TO[l]} \nonumber.
\ee
Here, the information about the graph cycle is retained but the information about the existence of another hypercycle is lost. CNs that only afford graph cycles are {\it complexed balanced}, otherwise they are called {\it deficient}. The deficiency of a network is precisely the number of basis hypercycles that are not cycles (that is, that are lost in passing to the representation in terms of complexes).} Around the concept of deficiency revolve many key results regarding the existence, uniqueness and stability of steady states\cite{feinberg,craciun06}.

An active research topic in biochemistry is the identification of basis vectors for the kernel and the cokernel of $\nabla$ affording a clear interpretation. We mention extreme pathways analysis \cite{palsson2,price}, which entails a classification of cycles according to their biochemical role.

On a complexed-balanced network, there is a standard procedure to identify a preferred basis of cycles affording a simple graphical interpretation, whose relevance to network thermodynamics has been explored by Schnakenberg for the stochastic description of master-equation systems\cite{schnak}. Interestingly, this construction identifies both the cycle vectors $\bs{c}_\gamma$ {\it and} their duals $\bs{c}^\dagger_\gamma$, bestowing on them a clear physical interpretation in the spirit of Kirchhoff's mesh analysis of electrical circuits. It involves the identification a basis of cycles crossing the branches of a spanning tree. For example, the CCN in Sec.\;\ref{example} has
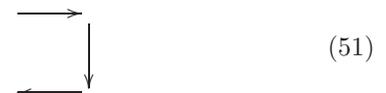
\be
\ba{c}\xymatrix{ \ar@{->}[r] & \ar@{->}[d] 
\\   & \ar@{->}[l] }\ea \ee
as spanning tree. Adding the edge $\bs{c}^\dagger_1$ that is left out identifies  the only cycle $\bs{c}_1$:
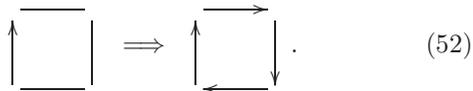
\be
\ba{c}\xymatrix{ \ar@{-}[r] \ar@{<-}[d]  & \ar@{-}[d] 
\\   & \ar@{-}[l] }\ea \implies \ba{c}\xymatrix{ \ar@{->}[r] \ar@{<-}[d]  & \ar@{->}[d] 
\\   & \ar@{->}[l] }\ea .
\ee
Similarly, a spanning tree for the OCN is
\be \ba{c}\xymatrix{ &  \ar@/_/[l]  \ar@/^/[r]  & }  \ea. \ee
Adding the remaining edges one obtains the two cycles depicted in Eq.\,(\ref{eq:cycless}). 
Whilst there exist related concepts applied to hypergraphs \cite{hyperflows}, we are not aware of an analogous graphical construction.

{\color{black} While we will not insist on the graphical methods, it is important to appreciate that a CN is generally not a graph. While this does not affect thermodynamics at the mean field level, it does have consequences as as regards the stochastic thermodynamics. We plan to analyze these aspects in a future publication.}

\section{\label{close}Closed Chemical Networks}

Finally, we are in the position to define a CCN. We base our approach not on an effective description of  a macroscopic network believed to describe some complex (bio)chemical mechanism, but rather on fundamental constraints posed by the ultimately collisional nature of elementary reaction processes. 

We define a CCN as a CN that satisfies the following physical requirements, which we discuss below: Mass conservation, all elementary and reversible reactions, detailed balance. We conclude this section by introducing thermodynamic potentials and entropic balance.

\subsection{Mass conservation}

Since in a closed box there is no net exchange of matter with the environment, we require the existence of a conservation law $\bs{\ell}_1 = \bs{m}$ with all  positive integer entries, corresponding to the molar masses of the individual species. The corresponding conserved quantity $M$ is the total mass of reactants. This assumption will play a role as regards the existence of a ``ground chemostat'' and for the correct counting of chemical affinities.

A fundamental conservation law bearing similar consequences is electric charge conservation when ionized chemicals and electrochemical potentials across membranes are included into the picture. This is crucial for the correct thermodynamic modeling of all oxidation-reduction reactions. While the treatment follows similarly as for mass conservation, for simplicity we will not explicitly deal with it here. 

\subsection{Reversible, elementary reactions}

By the principle of microscopic reversibility, any reaction run forward can in principle be reversed. In fact, unitarity of the quantum laws of interaction between molecules implies that collision events run backward have the exact same transition amplitude as forward ones. As per the Boltzmann equation, time asymmetry is entirely due to the interaction of molecules with the environment before and after a collision, which determines the statistics of boundary states. Such states are assumed to be Maxwellian, so that any initial state has a finite probability and the inverse rate never vanishes.

Irreversible reactions are often encountered in the literature, but they should be seen as effective descriptions of more complicated underlying sequences of reactions whose net effect is to make the occurrence of a backward process negligible compared to the net current. Besides being physical, reversibility is a quite convenient assumption as it allows to work with vector spaces, while much of the literature on CNs is often restrained to currents living in polyhedral cones with {\it ad hoc} choices of convex basis vectors at the boundary of these regions, e.g. extreme pathways \cite{palsson2}.

Collisions involving three or more molecules at once (that is, within the short time-scale of the interaction between molecules), while not impossible, are extremely rare. Multimolecular reactions are usually the net result of sequences of elementary (i.e., mono- or bi-molecular) reactions that have not been resolved. Since between one reaction and the successive thermalization interjects, thermodynamics is affected by coarse graining. In particular, non-elementary reactions might be thermodynamically not independent one of another (as they might share an intermediate step), resulting in nonnull off-diagonal Onsager coefficients. Already Alan Turing observed that the law of mass-action must only be applied to the actual reactions, and not to the final outcomes of a number of them\cite{turing}.

\subsection{\label{detbal}Detailed balance}

The connection between reversibility and cycles was already present in Boltzmann's derivation of the H-theorem, where the problem of closed cycles of collisions was considered\cite{tolman,cercignani}. In brief, detailed balance can be formulated as: The product of the rate constants around a cycle is equal to the product of the rate constants along the reversed cycle. This does {\it not} imply that forward and backward rate constants must be equal, but that time-asymmetric contributions to rate constants should all cancel out along closed cycles, that is, they must be encoded in a state function. This is indeed the case we sketched at the beginning of the previous paragraph. Therefore, the condition of detailed balance is a structural property of rate constants that follows directly from microreversibility of collisions.  

Following a line of reasoning rooted in the stochastic thermodynamics of nonequilibrium systems, let us define the thermodynamic force as
\be
\bs{F} = \ln \frac{\bs{k}_+}{\bs{k}_-}.
\ee
Notice that it needs to be defined in terms of the scaled rate constants for dimensional consistency. The force is an intensive quantity.

Rate constants are said to satisfy detailed balance when the force is conservative, viz. when it is the gradient of a potential $\bs{\phi} = (\phi_s)_s$\cite{foot8},
\be
\bs{F}  = - \overline{\nabla}  \bs{\phi}. \label{eq:detbal}
\ee
As a consequence, letting $\rho_1,\ldots, \rho_n$ be the reactions forming cycle $\bs{c}$, the circulation of the force along any such cycle vanishes
\be
\bs{c} \cdot \bs{F} = \ln \frac{k_{+\rho_1} \ldots k_{+ \rho_n}}{k_{-\rho_1} \ldots k_{-\rho_n}} = - \bs{c} \cdot \overline{\nabla}  \bs{\phi} = 0, \label{eq:kolmogorov}
\ee
given that $\nabla \bs{c} = 0$. Seeing cycles as ``curls'', this corresponds to the vanishing of the curl of a gradient. The converse is also usually true: If the curl of a force vanishes, then the force is a gradient. This is the case in the present context: If $\bs{c} \cdot \bs{F} = 0$ along all cycles  (known as Kolmogorov criterion), then detailed balance holds\cite{foot9}. This criterion expresses a constraint on the rate constants for every independent cycle of the network.

From the definition of equilibrium steady state Eq.\;(\ref{eq:equivel}) and the mass-action law Eq.\;(\ref{eq:adim}) follows
\be
0 =  \ln \frac{\bs{j}^{eq}_+}{\bs{j}^{eq}_-}= \ln \frac{\bs{J}^{eq}_+}{\bs{J}^{eq}_-}  = \bs{F} - \overline{\nabla}  \ln\bs{z}^{eq}_{\bs{L}},  \label{eq:deteq}
\ee
where we remind that $\bs{z}$ denotes molar fractions. Therefore, if a system admits an equilibrium then $\bs{F}$ satisfies detailed balance, with $\bs{\phi} = -  \ln \bs{z}^{eq}_{\bs{L}} $. Vice versa, if $\bs{F}$ satisfies detailed balance, then the steady state is an equilibrium. We emphasize that detailed balance is a property of the rate constants, while being at equilibrium is a property of steady state molar fractions; this subtle distinction will be important when in the following we will assume rate constants to satisfy an analogous {\it local detailed balance}, but equilibrium will not follow.

Since $\bs{F}$ is defined independently of the state of the system, then it does not depend on the conserved quantities. After Eq.\;(\ref{eq:deteq}), letting $\bs{z}^{eq}_{\bs{L}'}$ be the molar fractions of an equilibrium state compatible with a different set of conserved quantities, one obtains $
\overline{\nabla}  \ln  \bs{z}^{eq}_{\bs{L}} = \overline{\nabla}   \ln  \bs{z}^{eq}_{\bs{L}'}$. Then any two equilibrium molar fractions are separated by a conservation law in the following way 
\be
\ln \bs{z}^{eq}_{\bs{L}} = \ln \bs{z}^{eq}_{\bs{L}'} + \sum_\lambda \theta_\lambda \bs{\ell}_\lambda, \label{eq:indepl}
\ee
where $\theta_\lambda$ are suitable coefficients. This equation defines an equivalence class of equilibrium steady states, with generic element $\bs{z}^{eq}$ encompassing all possible equilibra of mass-action chemical systems.

As an example, consider the CCN in Eq.\;(\ref{eq:excn}). Since all currents vanish, the equilibrium molar fractions obey
\bea
k_{+1}  z_1z_2 & = & k_{-1} z_3  \nonumber \\
k_{-2} z_4 z_2 & = & k_{+2}  z_3 \nonumber \\
k_{+3}  z_4z_2 & = & k_{-3} z_5 \nonumber \\
k_{-4} z_1z_2 & = & k_{+4}  z_5 . \label{eq:sysex}
\eea
Dividing the first by the second and the third by the forth and multiplying yields the following Kolmogorov criterion for the rate constants
\be
\frac{k_{+1} k_{+2} k_{+3} k_{+4}}{k_{-1} k_{-2} k_{-3} k_{-4}} = 1. 
\ee
Also, one can easily verify that any two solutions of Eq.\;(\ref{eq:sysex}) satisfy
\be
\left( \ba{c} \ln z'_1/z_1 \\ \ln z'_2/z_2 \\ \ln z'_3/z_3 \\ \ln z'_4/z_4 \\ \ln z'_5/z_5     \ea \right)  = \theta_1 \left(\ba{c} 1 \\ 1 \\ 2 \\ 1 \\ 2 \ea\right) + \theta_2 \left(\ba{c} 1 \\ 0 \\ 1 \\ 1 \\ 1 \ea\right) \ee
with $\theta_1 =  \ln z'_2/z_2$ and $\theta_2 =  \ln z'_1z_2/z_1z'_2 $.

\subsection{Thermodynamic potentials}

Let us introduce the chemical potential per mole
\be
\bs{\mu} = \bs{\mu}_{\bs{L}}^0 + R T \ln \bs{z},
\ee
where $\bs{\mu}^0_{\bs{L}}$ are standard chemical potentials (at state $T = 298.15 \, \mathrm{K}$, $p = 1 \,\mathrm{bar}$ and fixed values $\bs{L}$) and $R$ is the gas constant. Again, notice that chemical potentials are better defined in terms of adimensional molar fractions rather than concentrations. The free energy increase along a reaction is the weighted difference between the chemical potentials of reactants and products,
\be
\Delta \bs{G} = \overline{\nabla} \bs{\mu} = \Delta \bs{G}^0 + RT\, \overline{\nabla}   \ln \bs{z}.
\ee
By definition of standard chemical potentials, equilibrium steady states are characterized by a vanishing free energy difference, yielding
\be
\Delta \bs{G}^0 = - R T\,\overline{\nabla} \ln \bs{z}^{eq}_{\bs{L}} \label{eq:stanfed}
\ee
and $\bs{z}^{eq}_{\bs{L}} \propto e^{-\bs{\mu}_{\bs{L}}^0/(RT)}$. We obtain
\be
\Delta \bs{G} = R T\, \overline{\nabla} \ln  \frac{\bs{z}}{\bs{z}^{eq}} = - R T\, \ln \frac{\bs{j}_+}{\bs{j}_-}
\ee
and one can express the force as $\bs{F} = - \Delta \bs{G}^0/(R T)$.

Finally, and most importantly for this paper, the EPR {\color{black} of the full network is  given by \cite{}}
\be
T\Sigma = - \bs{j} \cdot \Delta \bs{G} = RT \,(\bs{j}_+ - \bs{j}_-)  \cdot \ln \frac{\bs{j}_+}{\bs{j}_-} \geq 0. \label{eq:EPR}
\ee
{\color{black} This expression is usually formulated in the literature \cite{kondepudi} in terms of the {\it extent} of reaction $d\xi_\rho = j_\rho \, dt$.} By the latter identity the EPR is positive and it only vanishes at equilibrium. Use of the scaled currents, rather than the molar currents $\bs{J}$, yields the right physical dimensions of an entropy per mole per unit time.

Since $\bs{z}$ is positive and normalized, we can interpret it as a probability and introduce its information-theoretic Shannon entropy (per mole) $\mathcal{S} = - R \, \bs{z} \cdot \ln \bs{z}$. Taking the time derivative and after a few manipulations one obtains
\bea
T\Sigma = - \frac{d\mathcal{G}_{\bs{L}}}{dt} -  \dot{\mathcal{N}}  \mathcal{G}_{\bs{L}}, \label{eq:EPRCN}
\eea
where we used the total rate defined in Eq.\;(\ref{eq:contev}), and introduced the average molar free energy
\be
\mathcal{G}_{\bs{L}} = RT \, \left\langle \ln \frac{\bs{z}}{\bs{z}_{\bs{L}}^{eq}}\right\rangle = \left\langle \bs{\mu}^0_{\bs{L}} \right\rangle - T \mathcal{S}, \label{eq:avG}
\ee
where the mean $\langle \, \cdot \, \rangle$ is taken with respect to the molar fractions $\bs{z}$. The average free energy $\mathcal{G}_{\bs{L}}$ is a relative entropy, hence it is nonnegative and it only vanishes at equilibrium. Also, notice that while it depends on conserved quantities, the balance equation Eq.\,(\ref{eq:EPRCN}) does not.

The last term in Eq.\,(\ref{eq:EPRCN}) is peculiar. It emerges because the total number of reactants is not conserved. Physically, considering that in biophysical conditions of constant temperature and pressure an increase in the total concentration of the system (say, the cell) produces an increase in volume, then the term $\dot{\mathcal{N}} \mathcal{G}_{\bs{L}}$ might be identified as a mechanical work performed by the system on its surrounding, with an effective pressure that depends on the internal free energy content (per volume). This term can be reabsorbed by expressing the above equation in terms of extensive quantities $[Z]\Sigma$, with dimension of an entropy per volume, and $[Z]\mathcal{G}_{\bs{L}}$, yielding
\be
T [Z] \Sigma = - \frac{d}{dt} \left([Z] \mathcal{G}_{\bs{L}} \right).
\ee
Finally, the standard chemical potential can be further split into enthalpy of formation and internal entropy of the molecules, $\bs{\mu}^0_{\bs{L}} = \bs{H}^0_{\bs{L}} - T\bs{S}$, yielding the following expression for the overall network free energy 
\be
\mathcal{G}_{\bs{L}} = \big\langle \bs{H}^0_{\bs{L}} \big\rangle - T \left( \mathcal{S} + \big\langle \bs{S} \big\rangle\right).
\ee
Notice that both internal degrees of freedom of molecules and their distibution contribute to the total entropy.

\subsection{Thermodynamic network independence}

In this section we further delve into the important fact that in all expressions involving free energy differences, conserved quantities do not appear.

Indeed, all incremental quantities, $\Delta \bs{G}^0$, $\Delta \bs{G}$, and the EPR, do not depend on $\bs{L}$, as can be seen by plugging Eq.\,(\ref{eq:indepl}) into Eq.\,(\ref{eq:stanfed}). Indeed, if $\Delta \bs{G}^0$ depended on the initial composition of the mixture, it would not be standard. This crucial property, which we dub {\it thermodynamic network independence}, is a consequence of the {\it relative} definition of free energy increase as a stoichiometric difference and of the mass-action law, which is implied in Eq.\;(\ref{eq:indepl}). Instead, {\it absolute} values $\bs{\mu}^0_{\bs{L}}$, $\mathcal{G}_{\bs{L}}$ and $\bs{H}^0_{\bs{L}}$ do depend on conserved quantities. In general, it is agreed that a special representative $\bs{\mu}^0$ from the equivalence class of standard chemical potentials is chosen by fixing the free energy of individual elements (in their most stable compound) to zero\cite{voet}, and calculating standard chemical potential differences $\Delta \bs{G}^0$ as free energies of formation of one mole of substance out of its composing elements. In this latter picture, conservation laws are implicitly tuned so as pick the individual elements composing the substance of interest and none of the others, being elements the ultimate conserved quantities in chemistry in terms of which every other conservation law can be (in principle) understood.

Let us emphasize the importance of thermodynamic network independence for thermodynamic modeling. By switching conservation laws on and off one can tune which reactants take part to the reaction, determining the network structure.  Thermodynamic network independence grants that state functions associated to chemicals of observable interest are independent both of the path of formation {\it and} of which other substrates take part to the reaction and how they evolve. Completing Voet \& Voet's words \cite{voet} (p. 54): ``The change of enthalpy in any hypothetical reaction pathway can be determined from the enthalpy change in any other reaction pathway between the same reactants and products {\it independently of the transformations undergone by the other reactants and products taking part to the reaction}''. Importantly, network independence and being a state function are {\it not} the same thing. In fact, the latter is a consequence of the definition of standard free energy differences Eq.\,(\ref{eq:stanfed}). The former also invokes Eq.\,(\ref{eq:indepl}), which is a consequence of the mass-action law, and of the fact that $\bs{L}$ are conserved, which follows from Eq.\,(\ref{eq:cont}). All three these equations are in principle independent, and they involve the stoichiometric matrix under very different incarnations.

We could not find an explicit discussion of this issue, probably because it only emerges in networks with conserved quantities. A consequence (yet speculative) might be that the estimation of free energies of nonelementary metabolic processes that do not obey mass-action kinetics, as carried out by means of various group contribution methods\cite{group}, are not a priori granted to be network-independent.

\section{\label{open}Open chemical networks}

\subsection{Chemostats}

An OCN is obtained by fixing the concentrations of some external species, called {\it chemostats}\cite{foot10}. Letting $\bs{Z} = (\bs{X},\bs{Y})$, we distinguish $s_X$ internal species with variable concentration $[\bs{X}]$ and $s_Y$ chemostats with fixed concentration $[\bs{Y}]$, with $s_X + s_Y=s$. We identify internal and external stoichiometric sub-matrices $\nX, \nY$ obtained by eliminating the rows corresponding to the chemostats or the internal species, respectively.

For sake of ease, it is convenient to re-label species in such a way that $\overline{\nabla} = (\overline{\nabla}^X,\overline{\nabla}^Y)$. By definition, the kinetic equations for the OCN read
\be
\frac{d}{dt} [\bs{X}] = \nX \bs{J},  \qquad \frac{d}{dt}  [\bs{Y}] = 0 \label{eq:kineqop}
\ee
with net reaction flux
\be
\bs{J} = \bs{K}_+  \, [\bs{Y}] ^{\,\cdot\nY_{\!+}} [\bs{X}]^{\,\cdot\nX_{\!+}}   - \bs{K}_- \, [\bs{Y}]^{\,\cdot\nY_{\!-}} [\bs{X}]^{\,\cdot\nX_{\!-}}.
\ee
We further introduce the rates at which the environment provides chemostats to the system to maintain their concentrations as
\be
\frac{\dbar}{\dbar t} [\bs{Y}] = \nY \bs{J}. \label{eq:chemflux}
\ee

Inspecting the above expressions one realizes that OCN kinetics is equivalent to mass-action kinetics on the CN
\be
\xymatrix{
\nX_{\!+} \bs{X}  ~ \FROM[r]_{\tilde{\bs{K}}_-}\TO^{\tilde{\bs{K}}_+}[r] & ~ \nX_{\!-} \bs{X}}
\ee
with modified rate constants given by
\be
\tilde{\bs{K}}_{\pm} = \bs{K}_{\pm} \, [\bs{Y}]^{\,\cdot\nY_{\! \pm }}.
\ee
A complex whose species are all chemostatted is denoted $\emptyset$. With little loss of generality, we assume that no reaction with vanishing column $\nabla^X_{\!\rho} = 0$, contributing a null current, is obtained.

{\color{black}In our example, we have the following splitting
\be
\nX =  \left(\ba{cccc}  -1 & 1 & -1 & 1 \\ 1 & -1 & 0 & 0 \\  0 & 0 & 1 & -1 \ea\right) , \quad \nY = \left(\ba{cccc} -1 & 0 & 0 & 1 \\  0 & 1 & -1 & 0 \ea\right).
\ee
The reaction fluxes read
\bea
J_{1} & = & \tilde{K}_{+1} [X_2] - \tilde{K}_{-1} [X_3] \nonumber \\
J_{2} & = & \tilde{K}_{+2} [X_3] - \tilde{K}_{-2} [X_2] \nonumber \\
J_{3} & = & \tilde{K}_{+3} [X_2] - \tilde{K}_{-3} [X_5] \nonumber \\
J_{4} & = & \tilde{K}_{+4} [X_5] - \tilde{K}_{-4} [X_2] 
\eea
with modified rate constants
\bea
\tilde{K}_{+1} = K_{+1} [Y_1], &\qquad& \tilde{K}_{-1} = K_{-1} \\
\tilde{K}_{-2} = K_{-2} [Y_4], &\qquad& \tilde{K}_{+2} = K_{+2} \\
\tilde{K}_{+3} = K_{+3} [Y_4], &\qquad& \tilde{K}_{-3} = K_{-3} \\
\tilde{K}_{-4} = K_{-4} [Y_1], &\qquad& \tilde{K}_{+4} = K_{+4},
\eea
where the concentrations $[Y_1]$ and $[Y_4]$ held constant. The mass-action kinetic equations are given by
\bea \label{eq:exmakop}
\tfrac{d}{dt} [X_2] & = & J_{4} - J_{1} + J_2 - J_3 \nonumber \\
\tfrac{d}{dt} [X_3] & = & J_{1} - J_{2} \nonumber \\
\tfrac{d}{dt} [X_5] & = & J_{3} - J_{4}
\eea
and the rate of injection of the chemostat are given by 
\be
\frac{\dbar}{\dbar t} [Y_1] = J_{4} - J_{1} , \qquad
\frac{\dbar}{\dbar t} [Y_4] = J_{2} - J_{3} .
\ee
}

\subsection{\label{water}Solvents as the ground chemostats}

A delicate issue was left aside. Mass-action kinetics accounts for variations of the concentrations only due to changes of the particle number in a given constant volume $V$. Biochemical processes are typically both isothermal and isobaric\cite{voet}. Then, volume cannot be constant in general, its variation depending on the equation of state of the mixture relating $V,T,p,[\bs{Z}]$. The latter is usually unknown, and often ideal behavior, as of dilute gases, is assumed. Then, Eq.\;(\ref{eq:cont}) is only valid insofar as processes are approximately isochoric. While this is often the case, it must be recognized that this is an approximation.

Constant volume occurs when the mixture is a dilute solution, i.e. a solvent $Z_1$ occupies most of the volume, and $[Z] \approx [Z_1]$. Since the standard solvent for biochemical processes is water, one might want to identify $[Z_1] = [\WATER]/55.5 = 1 \, \mathrm{mole}/\mathrm{liter}$, in which case we can simply replace $[Z]=1$ and $[\bs{Z}]$ by the molar fractions $\bs{z}$. 

The solvent might be active or passive according to whether it partakes to chemical reactions or not. In the latter case, in the CCN $d [Z_1]/dt = 0$ and taking $[Z_1] \gg [Z_{\sigma>1}]$ permits to write
\be
\frac{d \bs{z}}{dt} \approx \nabla \bs{j}. \label{eq:molfrac}
\ee
Under these assumptions the total rate $\dot{\mathcal{N}} \approx 0$ and one recovers an expression of the EPR in Eq.\;(\ref{eq:EPRCN}) as time derivative of the average free energy. Notice that a passive solvent contributes a null column to the stoichiometric matrix, so it falls outside of our assumptions.

In the former case of an active solvent, we regard it as the first chemostat. Imposing $d [Y_1]/dt = 0$ and assuming $[Y_1] \gg [Z_{\sigma>1}]$ we obtain
\be
\frac{d \bs{x}}{dt} \approx \nX \bs{j},  \qquad \frac{d \bs{y}}{dt} \approx 0 \qquad \mathrm{etc.} \label{eq:kineqop}
\ee

We will attain to this scenario in the following: Certain species are approximately constant; At least one of them is a solvent, i.e. its concentration is overwhelmingly larger than the others. Relaxing this assumption results in very subtle issues, because one should then couple mass-action kinetics of reactions in local volumes to viable global equations of state to obtain more general dynamics. This interesting problem has not yet been addressed, to our knowledge.

\subsection{Emergent cycles}

Imposing some species to become chemostats might lead to emergence of new cycles whenever two previously distinguished complexes become undistinguishable, e.g. $Y_1 + X_2 \sim Y_4+X_2$, or more generally whenever a new pathway from a set reactants to a set of products becomes feasible (on an hypergraph this might not be easily visualizable). Correspondingly, the number of KCLs to be enforced diminishes, making the span of steady currents larger. In fact, cycles $\bs{c} \in \mathscr{C}$ of the closed network also trivially belong to the kernel $\tilde{\mathscr{C}}$ of $\nX$, since
\be
\left(\ba{c} \nabla^X \\ \nabla^Y \ea\right) \bs{c} = \left(\ba{c} \nabla^X \bs{c} \\ \nabla^Y \bs{c} \ea\right) = 0.
\ee
The reverse is not true, since there might exist vectors $\bs{c}_\alpha$ such that
\be
\left(\ba{c} \nabla^X \\ \nabla^Y \ea\right) \bs{c}_\alpha = \left(\ba{c} 0 \\ \nabla^Y \bs{c}_\alpha \ea\right) \neq 0.
\ee
{\color{black} In this case we talk of {\it emergent} cycles. 
We label by index $\alpha$ a set $\bs{c}_\alpha$, such that $\nabla^X \bs{c}_\alpha = 0$, $\nabla \bs{c}_\alpha \neq 0$ and such that they are linearly independent among themselves and with respect to the closed network cycles $\bs{c}_\gamma$}. It follows that a steady configuration of currents satisfying $\nabla^X \bs{j}^\ast = 0$ has general solution
\be
\bs{j}^\ast = \sum_\gamma \mathcal{J}_\gamma \bs{c}_\gamma + \sum_\alpha \mathcal{J}_\alpha \bs{c}_\alpha, \label{eq:cyclecurr}
\ee
where $\gamma$ spans cycles of the CCN and $\alpha$ spans the emergent cycles.

{\color{black}
In our example, the two cycles $\overline{\bs{c}}' = (0,0,1,1)$ and $\overline{\bs{c}}'' = (1,1,0,0)$ are emergent. Only one of the two is independent of the other and of the closed network cycle $\overline{\bs{c}} = (1,1,1,1)$. It follows that the steady state solution of the Eqs.\,(\ref{eq:exmakop}) lives in the current vector space spanned by $\bs{c}$ and $\bs{c}'$.}

% In particular that if we plug in the equilibrium values that the variable species would have attained in the closed network we obtain nonvanishing steady currents, unless also the chemostats' concentrations are tuned to their equilibrium value. 

\subsection{Local detailed balance and affinities}

The thermodynamic force is now defined in terms of the modified rates
\be
\tilde{\bs{F}} := \ln \frac{\tilde{\bs{k}}_+}{\tilde{\bs{k}}_-} = \bs{F} - \onY \ln \bs{y}.
\ee
Splitting the chemical potential $\bs{\mu} = (\bs{\mu}^X,\bs{\mu}^Y)$, choosing one preferred representative $\bs{\mu}^0$ in the equivalence class of equilibrium standar chemical potentials, and introducing the internal free energy differences
\be
\Delta \bs{G}^{X}_{\bs{L}} =  \overline{\nabla}^{X} \bs{\mu}^{X} = RT \, \overline{\nabla}^{X} \ln \frac{\bs{x}}{\bs{x}^{eq}_{\bs{L}}}\qquad \text{(resp. $Y$)} \label{eq:deltag} % \\
% \Delta \bs{G}^{Y} = \overline{\nabla}^{Y} \bs{\mu}^{Y} = \overline{\nabla}^{Y} \ln \frac{\bs{y}}{\bs{y}^{eq}},
\ee
we can express the thermodynamic force as
%\be
%\tilde{\bs{f}} =  \onX \left(  \bs{\mu}^{X,eq,\bs{L}} - \bs{\mu}^X_0 \right) + \onY \left( \bs{\mu}^{Y,eq,\bs{L}} - \bs{\mu}^Y \right). \label{eq:locdetbal}
%\ee
\be
\tilde{\bs{F}} % = - \nicefrac{1}{RT} \left( \onX \bs{\mu}^{X,0} - \onY \bs{\mu}^Y\right)
 = - \frac{1}{RT} \left( \Delta \bs{G}^{0,X}_{\bs{L}} + \Delta \bs{G}^Y_{\bs{L}} \right). \label{eq:locdetbal}
\ee
We will refer to Eq.\,(\ref{eq:locdetbal}) as the condition of {\it local detailed balance} \cite{esposito1,maes1}. It must be emphasized that in this case the force is {\it not} a gradient force, as strict detailed balance is satisfied when there exists a potential $\tilde{\bs{\phi}}$ such that $\tilde{\bs{F}} = - \overline{\nabla}^X \tilde{\bs{\phi}}$. The physical picture is that of several reservoirs competing to impose their own rule and leading to a nonequilibrium state. In this approach local detailed balance does not require the baths to be at an equilibrium state. It suffices that certain resources are available in a controlled way. In fact, mechanisms that provide such resources to metabolic networks (respiration, radiation, nutrition) are  themselves nonequilibrium processes.

The circulation of the force along cycles of the closed network vanishes,
\be
\bs{c}_\gamma \cdot \tilde{\bs{F}} =  - \bs{c}_\gamma \cdot \onY \ln \bs{y}    = 0,
\ee
where we employed Kolmogorov's criterion Eq.\,(\ref{eq:kolmogorov}). These are thermodynamically reversible cycles. The circulation of the force along emergent cycles yields nonnull (De Donder\cite{dedonder,prigo,nicolis}) {\it affinities}
\be
\mathcal{A}_\alpha = \bs{c}_\alpha \cdot \tilde{\bs{F}} =  - \bs{c}_\alpha \cdot\onY \ln \frac{\bs{y}}{\bs{y}^{eq}_{\bs{L}}}  = - \frac{\bs{c}_\alpha \cdot \Delta \bs{G}^Y_{\bs{L}}}{R T} . \label{eq:affinity}
\ee
Affinities are particular linear combinations of the chemical potential differences between chemostats. They are the fundamental forces describing the steady state properties of an OCN. Dependence on standard chemical potentials, which are unphysical reference values, has disappeared. It will be a consequence of Sec.\,\ref{conservation} that affinities do not depend on $\bs{L}$ either, while in general $\Delta G^{X}_{\bs{L}}$ and $\Delta G^{Y}_{\bs{L}}$ do.

The emergence of nonvanishing affinities is due to the fact that there does not exist a free energy landscape for the OCN. This occurs because certain states are lumped, and the free energy is no longer single-valued at these states. The free energy increase depends on the path between lumped states, and its circuitation along emergent cycles does not vanish. Strict detailed balance is satisfied if and only if all affinities vanish, that is when the chemostats' chemical potentials $\bs{\mu}^Y$ attain their equilibrium values up to a conservation law, as we will discuss below. Notice nevertheless that there is a role to free energy differences in OCN, both for the definition of the thermodynamic force and the expression of affinities.

\subsection{Entropy production rate}

We define the internal and external EPRs as
\be
T \Sigma^{X,Y}_{\bs{L}} = - \bs{j} \cdot \Delta\bs{G}^{X,Y}_{\bs{L}},
\ee
while the total EPR is defined as in Eq.\;(\ref{eq:EPR}). In terms of the modified rates it reads
\be
T\Sigma = RT\, \sum_\rho \left( \tilde{k}_{+\rho} \, \bs{x}^{\,\cdot\nabla_{\!+\rho}} - \tilde{k}_{-\rho} \, \bs{x}^{\,\cdot\nabla_{\!-\rho}}  \right)\ln \frac{ \tilde{k}_{+\rho} \, \bs{x}^{\,\cdot\nabla_{\!+\rho}}}{\tilde{k}_{-\rho} \, \bs{x}^{\,\cdot\nabla_{\!-\rho}} }.
\ee
The total EPR is positive, encoding the second law of thermodynamics, and it vanishes at equilibrium. Let us notice in passing that the latter expression is usually postulated for the thermodynamic description of master-equation systems\cite{schnak} and it exactly coincides with it when the network is linear.  In particular, in the spirit of Prigogine's entropic balances \cite{prigo}, one has a natural identification of internal entropy production and of fluxes to the environment as follows.

In analogy to Eq.\;(\ref{eq:avG}) we define an average free energy difference for the internal species
\be
\mathcal{G}^X_{\bs{L}} = \hat{\bs{x}} \cdot \bs{\mu}^{0,X}_{\bs{L}} - T \mathcal{S}^X  .
\ee
where letting $x = \bs{1} \cdot \bs{x}$ be the total molar fraction of the internal species, $\hat{\bs{x}} = \bs{x}/x$ is the probability of an internal molecule and $\mathcal{S}^X = - R\,\hat{\bs{x}} \ln \hat{\bs{x}}$ is its entropy. Using the kinetic equations Eq.\,(\ref{eq:kineqop}) and Eq.\;(\ref{eq:deltag}), after some manipulations one obtains for the internal EPR
\be
T \Sigma^X_{\bs{L}} = -  \frac{d}{dt} [x \, \mathcal{G}^X_{\bs{L}} + RT( x \ln x - x)]. \label{eq:gibbseq}
\ee
Following from the fact that $\Delta \bs{G}^X_{\bs{L}}$ is a state function, the internal EPR is a total time derivative. {\color{black} Interestingly, there is a volume contribution to the internal entropy balance due to the fact that the total molar fraction of internal species is not conserved.}

As regards the external entropy balance, from the definition of chemostat current Eq.\.(\ref{eq:chemflux}) we obtain
\be
T\Sigma^Y_{\bs{L}} =  - \bs{\mu}^{Y} \cdot \frac{\dbar \bs{y}}{\dbar t},
\ee
This is not a total time derivative. Interestingly, the external EPR can be expressed solely in terms of chemostats' chemical potentials and fluxes.

\subsection{Cyle decomposition of steady EPR}

At a steady state the internal EPR vanishes and we obtain
\be
T\Sigma^\ast = - \bs{\mu}^{Y} \cdot \left( \frac{\dbar \bs{y}}{\dbar t} \right)^\ast. \label{eq:eprchem}
\ee
This elegant formula allows to compute the total dissipation of a CN solely in terms of observables associated to its chemostats, without knowing details about the internal species. The internal structure of the CN does play a role as regards the response of the currents to perturbations of chemical potentials.
 
We can further compress the expression of the steady EPR using Eq.\,(\ref{eq:chemflux}) and the the steady state solution in terms of the cycle currents, Eq.\;(\ref{eq:cyclecurr}):
\bea
T\Sigma^\ast & = & -  \sum_{\gamma} \mathcal{J}_{\gamma} \, \bs{\mu}^{Y} \! \cdot \! \nY  \bs{c}_\gamma -  \sum_{\alpha} \mathcal{J}_{\alpha}  \, \bs{\mu}^{Y} \! \cdot \! \nY  \bs{c}_\alpha \nonumber \\
& = & RT \sum_\alpha \mathcal{J}_\alpha \mathcal{A}_\alpha. \label{eq:schnakenberg}   
\eea
This fundamental expression is the CN analog of a celebrated result by Hill and Schnakenberg\cite{hillfree,schnak}. Importantly, not all cycles actually contribute to the steady EPR, but only those that were originated after the network collapse. It suffices that all $\mathcal{J}_\alpha = 0$ to make the EPR vanish. On the other hand we know that all currents need to vanish at the equilibrium steady state. Then $\mathcal{J}_\alpha = 0$ implies $\mathcal{J}_\gamma = 0$. So, cycle currents are not independent one of another. Indeed there is a high degree of correlation of steady currents internal to a CN. We will analyze consequences on the linear response and on network reconstruction in the follow-up paper.

Also, observe that in general the number of affinities is less than the number of chemostats (see Sec.\ref{number}), so that Eq.\;(\ref{eq:eprchem}) is redundant. Eq.\;(\ref{eq:schnakenberg}) is the most compressed and fundamental form for the steady EPR.

\subsection{Broken conservation laws and symmetries\label{conservation}}

In this paragraph we study the fate of conservation laws. The general understanding is that by providing fluxes of chemostats one might break the conservation of internal chemical species. As a trivial example, an open reactor initially empty will soon be populated. The second insight is that the relics of a broken conservation law manifest themselves as symmetries of the affinities and as linear relations between steady chemostat fluxes. 

As usual, we distinguish internal and external parts of a conservation law $\bs{\ell} = (\bs{\ell}^X,\bs{\ell}^Y)$. The balance of conserved quantities across the system's boundary can then be written as
\be 
\bs{\ell}^X \cdot \frac{d\bs{x}}{dt} + \bs{\ell}^Y \cdot  \frac{\dbar \bs{y}}{\dbar t} = 0, 
\ee
following from
\be
\overline{\left(\ba{c} \nabla^X \\ \nabla^Y \ea \right)} \left(\ba{c} \bs{\ell}^X \\ \bs{\ell}^Y \ea \right) = \onX \bs{\ell}^X  + \onY \bs{\ell}^Y = 0.
\ee
Notice that if $\bs{\ell}^X$ is a conservation law for $\nX$, then $\bs{\ell}^Y$ is a conservation law for $\nY$ and $\bs{\ell}$ is a conservation law of $\nabla$. The converse is not always true. Then, the number of conservation laws can only decrease. In particular, since all species have nonvanishing mass, by construction the mass conservation law $\bs{m}$ is always broken within the system as the first chemostat $Y_1$ is fixed, implying the mass balance across the system's boundaries
\be
\sum_{\sigma \neq 1} m_\sigma \nabla_{\sigma,\rho} = - m_1 \nabla_{1,\rho} \neq 0. \label{eq:massbalance}
\ee
Using the kinetic equation (\ref{eq:kineqop}) we obtain
\be
\bs{\ell}^X \cdot \dot{\bs{x}} = \bs{\ell}^X  \cdot \nX \bs{j} =  -\bs{\ell}^Y  \cdot \nY \bs{j} =  - \bs{\ell}^Y  \cdot \frac{\dbar \bs{y}}{\dbar t}. \label{eq:tautology}
\ee
If $\bs{\ell}^X$ is a conservation law for $\nX$ then $\bs{\ell}^Y \cdot \nY = 0$ and the above equation says that the quantity $L^X = \bs{\ell}^X \cdot \bs{x}$ is conserved. Now consider all conservation laws $\bs{\ell}_\beta$ that are broken, that is such that $\bs{\ell}^X_\beta \cdot \nX \neq 0$. The left hand side of Eq.\,(\ref{eq:tautology}) vanishes at the steady state, and we are left with
\be
\bs{\ell}_\beta^Y \cdot \left(\frac{\dbar \bs{y}}{\dbar t}\right)^\ast = 0,
\ee
that is, every broken conservation law gives an independent linear constraint on the steady chemostat fluxes. From now on $\beta$ will label a basis of $b$ independent broken conservation laws. In particular, given the mass balance Eq.\,(\ref{eq:massbalance}), we get
\be
\sum_{i > 1} m^Y_i  \left( \frac{\dbar  y_i }{\dbar t}\right)^\ast = - m^Y_1 \left( \frac{\dbar  y_1 }{\dbar t}\right)^\ast .
\ee
where $Y_1$ is the ground chemostat. Then, the expression Eq.\;(\ref{eq:eprchem}) of the EPR in terms of chemostats can be further compressed
\be
T\Sigma^\ast = - \sum_{i >1} \mu^{Y}_i  \left( \frac{\dbar  y_i }{\dbar t}\right)^\ast - \mu^{Y}_1  \left( \frac{\dbar  y_1 }{\dbar t}\right)^\ast = - \sum_{i >1} \tilde{\mu}_i^Y  \left( \frac{\dbar  y_i }{\dbar t}\right)^\ast
\ee
where
\be
\tilde{\mu}_i^Y =  \sum_{i >1} \left( \mu^{Y}_i -   \frac{m_i}{m_1} \mu^{Y}_1 \right).
\ee
Importantly, if there is only one chemostat, then the steady EPR vanishes. It takes at least two chemostats to generate a nonequilibrium current, because of mass conservation.

Broken conservation laws also play an important role as regards the conditions under which detailed balance is satisfied. Let us consider the behavior of affinities when the chemostat's chemical potentials are shifted by a linear combination of conservation laws,
\be
\delta \bs{\mu}^Y = \sum_\lambda \theta_\lambda \bs{\ell}^Y_\lambda. \label{eq:transchem}
\ee
After Eq.\,(\ref{eq:affinity}), we obtain
\bea
\delta \mathcal{A}_\alpha & = & - \frac{1}{R T} \sum_\lambda \theta_\lambda \bs{\ell}^Y_\lambda \cdot \nY \bs{c}_\alpha \nonumber \\ 
& = & + \frac{1}{R T} \sum_\lambda \theta_\lambda \bs{\ell}^X_\lambda \cdot \nX \bs{c}_\alpha  ~= ~ 0 
\eea
where the latter identity follows from $\nabla^X\bs{c}_\alpha = 0$. 
% For vanishing affinities, this implies that all concentrations of chemostats such that
%\be \frac{\bs{y}}{\bs{y}^{eq} } = \exp \frac{1}{R T}\sum_\lambda \theta_\lambda \bs{\ell}^Y_{\lambda}\ee
%maintain detailed balance and $\Sigma$ invariant.
Notice that being vectors $\bs{\ell}^Y_\lambda$ portions of full conservation laws, most of them will not actually be independent. In fact, it will turn out from the next section that there are $b$ independent symmetries of the affinities, i.e. independent transformations of the chemical potentials for which the affinities do not vary. It is tempting to notice that this subtle interplay between symmetries and conservations is somewhat reminiscent of the Noether paradigm in classical and quantum field theory. This also proves that affinities and the total EPR are independent of the reference values $\bs{L}$ chosen as standard state, while the internal and the external EPRs are not independent, as after Eq.\,(\ref{eq:transchem}) they transform according to
\be
\delta \Sigma_{\bs{L}}^{X} = - \delta \Sigma_{\bs{L}}^{Y} = - \frac{1}{T} \frac{d}{dt} \sum_\beta \theta_\beta L_\beta^X
\ee
where $L_\beta^X = \bs{\ell}_\beta^X \cdot \bs{x}$. That is, the definition of internal and external entropy production rates depends on the choice of reference state $\bs{\mu}^0_{\bs{L}}$ (see Ref.\cite{polettinigauge} for a similar analysis applied to master equation thermodynamics). 

To conclude, we mention that the distinction between internal and external conservation laws is also relevant to the classification of biochemical metabolic pools in the context of FBA\cite{palsson1}.

\subsection{Number of affinities and symmetries\label{number}}

{\color{black}We now turn to our most important finding.} 
Under chemostatting the number of cycles cannot decrease and the number of conservation laws cannot increase. Moreover, mass conservation is broken as the first chemostat is introduced. Consider the rank-nullity theorem Eq.\,(\ref{eq:ranknull}). When $s$ is decreased by one, at fixed $r$ (we are supposing no reactions are externalized), either $c$ increases by one or $\ell$ decreases by one. Then, the number $a$ of independent affinities and the number $b$ of broken conservation laws satisfy
\be
a+b = s^Y,
\ee
where we remind that $s^Y$ is the number of chemostats. This also proves that, since $a$ affinities are given in terms of $s^Y$ chemostats, then there must be $b$ independent ``Noether'' symmetries of the affinities.

Mass conservation implies $b \geq 1$ and 
\be
a \leq s^Y - 1.
\ee
That is, if a mass conservation law exists then there exists one ground chemostat relative to which all other chemostats are confronted. It takes at least two chemostats to drive the system out of equilibrium. As we discussed above, in physiological solutions water can be thought of as the ground chemostat. 

Notice that, while the mass conservation law is broken by definition, there might still survive conservation laws with all nonvanishing entries. The latter are important in that they grant the existence of a unique steady state \cite{fleming}.

\section{\label{conclu}Conclusions and perspectives}

In this paper we posed the foundations for the thermodynamic description of  OCNs subjected to influxes of species of fixed concentrations, the chemostats, a theme that recurrently captured the attention of researchers\cite{oster,clarke,qian}. The major novelties of our approach are the network characterization of thermodynamic observables, and the ``bottom-up'' approach grounded on the physical requirements that elementary reactions must obey. One straightforward takeaway is the relationship $s^Y = a + b$ between the number of chemostats, that of thermodynamics forces (the affinities) and that of conserved quantities across the system's boundary (broken conservation laws). Beyond this relation, there lie our main results: The entropy production rate can be expressed in terms of affinities and cycle currents, the former being given in terms of the chemostats' chemical potentials; the description can be further reduced by the effect of symmetries, entailed by broken conservation laws.

While we postpone a broad-scope discussion to the companion paper\cite{paper2}, let us here briefly comment on specific open questions that might deserve further attention in the future.

It is known that most metabolic pathways are close to equilibrium, and that only a few (e.g. those responsible for ADP/ATP conversion) are markedly out of equilibrium. Moreover, while there can be numerous internal species, there typically are few chemostats partaking to several reactions. As a consequence, only a few global thermodynamic cycles should be relevant for the complete thermodynamic modeling of metabolic networks, with locally equilibrated subnetworks feasible of coarse graining \cite{esposito3,altaner}. 

As regards the analysis of steady currents, an open question is whether there is an hypergraph procedure analogous to the spanning tree analysis to obtain a meaningful basis of cycles and of dual generating edges. Similarly, one could ponder whether there exists an analogous construction for a basis of conservation laws, and the relationship to extreme pathways.

OCNs are a subset of all possible chemical reaction networks to which, for example, deficiency theory applies. Some lines of inquiry are dedicated to the smallest CNs that present interesting behavior such as multiple steady states, Hopf bifurcations, periodicity, attractors, etc. As is the case when assuming elementary reactions \cite{wilhelm}, reversibility, detailed balance and the particular structure of emergent cycles/broken conservation laws might pose further constraints on this effort.

While intensive thermodynamic forces depend on relative concentrations independently of the total volume of the system, 
mass-action kinetics depends on actual concentrations. We believe that this subtle difference, so far unappreciated, might lead to interesting phenomena. 

Finally, the theory lends itself very naturally to generalization to periodic external driving\cite{markus}, e.g. through respiration and nutrition, underpinning circadian rhythms.

\SkipTocEntry \section*{Acknowledgments}

This work is dedicated to the memory of T. Hill (1917-2014), whose insights on cycle kinetics and free energy transduction of are of fundamental importance for the thermodynamic modelling of molecular processes.

The authors are grateful to H. Qian, R. Fleming and N. Vlassis for illuminating discussion and suggestions. The research was supported by the National Research Fund Luxembourg in the frame of project FNR/A11/02 and of the AFR Postdoc Grant 5856127.

\appendix

{\color{black}

\section{Example\label{appex}}

Consider the following CCN
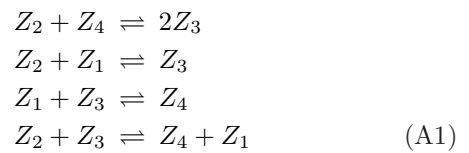
\bea
	Z_2 + Z_4  &~\mathop{\rightleftharpoons}~& 2Z_3 \nonumber \\
	Z_2 + Z_1  &~\mathop{\rightleftharpoons}~& Z_3 \nonumber \\
	Z_1 + Z_3  &~\mathop{\rightleftharpoons}~& Z_4 \nonumber \\
	Z_2 + Z_3 &~\mathop{\rightleftharpoons}~& Z_4 + Z_1 
\eea
with stoichiometric matrix
\be
\nabla = \left(\ba{cccc} 0 & -1 & -1 & 1 \\ -1 & -1 & 0 & -1 \\
2 & 1 & -1 & -1 \\ -1 & 0 & 1 & 1 \ea \right). \label{eq:stexap}
\ee
It affords a mass conservation law and a cycle
\be
\bs{m} = \left(\ba{c} 1 \\ 2 \\ 3 \\ 4 \ea \right), \qquad
\bs{c} = \left(\ba{c} 1 \\ -1 \\ 1 \\ 0 \ea \right),
\ee
which are respectively the left and right null vectors of the stoichiometric matrix. Notice that the cycle is a genuine hypercycle, since it does not preserve the complexes. The requirement that the network is closed, hence that detailed balance should be satisfied, implies the following condition on the rate constants
\be
\ln \frac{k_{+1} k_{-2} k_{+3}}{k_{-1} k_{+2} k_{-3}} = 0.
\ee
With this condition the mass-action kinetic equations (that we omit) admit a family of equilibria parametrized by the total mass $m = z_1+2z_2+3z_3+4z_4$. For example, from the last two reactions we find the equilibrium relation
\be
z^{eq}_2 = \frac{k_{-4}k_{+3}}{k_{+4}k_{-3}} (z^{eq})^2_1 \label{eq:zeq}
\ee
and similarly for the other concentrations.

We now chemostat species $Z_2$. We obtain
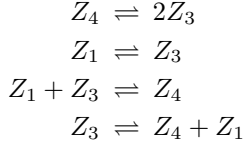
\bea
Z_4  &~\mathop{\rightleftharpoons}~& 2Z_3 \nonumber \\
Z_1  &~\mathop{\rightleftharpoons}~& Z_3 \nonumber \\
Z_1 + Z_3  &~\mathop{\rightleftharpoons}~& Z_4 \nonumber \\
Z_3 &~\mathop{\rightleftharpoons}~& Z_4 + Z_1 \nonumber
\eea
with stoichiometric matrix obtained by eliminating the second row in Eq.\,(\ref{eq:stexap}). The latter can be easily seen to afford one cycle and no conservation law. Hence the introduction of the fist chemostat broke mass conservation but did not break detailed balance. We further process chemostsatting species $Z_1$, to obtain the OCN
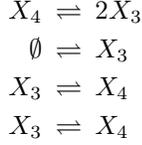
\bea
	X_4  &~ \mathop{\rightleftharpoons} ~& 2X_3 \nonumber \\
	\emptyset &~ \mathop{\rightleftharpoons} ~& X_3 \nonumber \\
	X_3  &~ \mathop{\rightleftharpoons} ~& X_4 \nonumber \\
	X_3 &~ \mathop{\rightleftharpoons} ~& X_4 \nonumber
\eea
with stoichiometric matrix
\be
\nabla^X = \left(\ba{cccc}
2 & 1 & -1 & -1 \\ -1 & 0 & 1 & 1 \ea \right).
\ee
and effective rates
\bea
\tilde{k}_{+1} = k_{+1}y_2, & \qquad &
\tilde{k}_{-1} = k_{-1} \nonumber \\
\tilde{k}_{+2} = k_{+2}y_1 y_2, & \qquad &
\tilde{k}_{-2} = k_{-2} \nonumber \\
\tilde{k}_{+3} = k_{+3}y_1, & \qquad &
\tilde{k}_{-3} = k_{-3} \nonumber \\
\tilde{k}_{+4} = k_{+4}y_2, & \qquad &
\tilde{k}_{-4} = k_{-4}y_1
\eea
The OCN affords the following emergent cycle, 
\be
\bs{c}' =  \left(\ba{c} 0 \\ 0 \\ 1 \\ -1 \ea \right)
\ee
carrying an affinity
\bea
\mathcal{A}(\bs{c}') & = & \ln \frac{\tilde{k}_{+3}\tilde{k}_{-4}}{\tilde{k}_{-3}\tilde{k}_{+4}} \,=\,  \ln \frac{k_{+3}k_{-4}}{k_{-3}k_{+4}} \frac{y_1^2}{y_2} \nonumber \\
& = & \ln \left(\frac{y_1}{y_1^{eq.}} \right)^2
	\left(\frac{y_2^{eq.}}{y_2} \right),
\eea
where we used Eq.\,(\ref{eq:zeq}) in the last passage. The mass-action kinetic equations for the OCN read
\bea
\tfrac{d}{dt} x_4 & = & - (\tilde{k}_{+1} + \tilde{k}_{-3} + \tilde{k}_{-4} ) x_4 \nonumber \\
& &  + (\tilde{k}_{+3} + \tilde{k}_{+4})x_3 + \tilde{k}_{-1} x_3^2 \nonumber \\
\tfrac{d}{dt} x_3 & = & 
\tilde{k}_{+2} + (2\tilde{k}_{+1}  + \tilde{k}_{-3} + \tilde{k}_{-4} ) x_4  \nonumber \\
& & - (\tilde{k}_{-2} + \tilde{k}_{+3} + \tilde{k}_{+4})x_3  - 2\tilde{k}_{-1} x_3^2 
\eea
The steady state equation is easily seen to lead to a quadratic equation, that can be easily solved. Interestingly, for certain values of the rate constants this model displays bistability.

The stationary cycle current conjugate to the above affinity is the one that flows along the forth reaction,
\be
\mathcal{J}(\bs{c}') = j_4^\ast = k_{+4} x_3^\ast - k_{-4} x_4^\ast
\ee
so that the steady entropy production rate reads
\be
\Sigma^\ast = R \, \mathcal{J}(\bs{c}') \mathcal{A}(\bs{c}')
\ee
This expression can also in principle be derived by plugging the exact solution of the mass-action kinetic equations into the general definition of the entropy production rate, which is a tedious calculation.

Finally, by definition the chemostat's currents read
\bes
\frac{\dbar y^\ast_1}{\dbar t} & = & - j^\ast_2 - j^\ast_3 + j_4  = 2 \mathcal{J}(\bs{c}') \\ 
\frac{\dbar y^\ast_2}{\dbar t} & = & - j^\ast_1 - j^\ast_2 - j_4 =  - \mathcal{J}(\bs{c}')
\ees
where we used $\nabla^X \bs{j}^\ast = 0$ leading to $j^\ast_1 + j^\ast_2 = 0$ and $j^\ast_2 + j^\ast_3 = -j^\ast_4$. Then it is easily verified that
\be
\Sigma^\ast = \frac{\dbar y^\ast_1}{\dbar t} \mu_1 + \frac{\dbar y^\ast_2}{\dbar t} \mu_2.
\ee
 
}


\begin{thebibliography}{22}
\bibitem{foot1} These creation/annihilation reactions are unphysical at a fundamental level but in an open CN they simply mean that all species but $Z$ are chemostats.
\bibitem{foot2} Notice that molecules with the same molecular mass can occur with different conformation, ionization etc.
\bibitem{foot4} Our modern understanding of this fact is resumed in the so-called fluctuation theorem $\mathcal{J}(\bs{c})/\mathcal{J}(-\bs{c}) = e^{\mathcal{A}(\bs{c}) } = 1$. An early formulation of this relation can be found in Hill\cite{hillfree}, Eq.\;(7.8).
\bibitem{foot5} Although, when kinetic parameters are unknown and one needs to infer the network's flows from knowledge of its topology, such constraints have to be taken into account\cite{qian}.
\bibitem{foot7} Loosening this assumption in first approximation only requires to include phenomenological activity coefficients with straightforward modifications of our results.
\bibitem{foot8} While $\nabla$ acts like a divergence, $\overline{\nabla}$ acts like a gradient, compatibly with the fact that the gradient of a scalar is a vector.
\bibitem{foot9} Proof: Since $\bs{c} \cdot \bs{F} = 0$ then $\bs{F}$ lives in the space orthogonal to all cycles, which is the row-space of $\nabla$. Then, $\bs{F}$ is a linear combination of the rows of $\nabla$, as expressed by Eq.\;(\ref{eq:detbal}).
\bibitem{foot11} This is also implicit in the usual definition $\mathrm{pH} = - \ln [\mathrm{H}^+]$, which is otherwise dimensionally inconsistent.
\end{thebibliography}
\end{document}